\documentclass[final,5p,10pt,twocolumn]{elsarticle}
\biboptions{numbers,sort&compress} 

\usepackage[utf8]{inputenc}
\usepackage{setspace}
\usepackage{epsfig}    
\usepackage{amsfonts}
\usepackage{amssymb}
\usepackage{parskip}

\usepackage{amsmath}
\usepackage{comment}
\usepackage{multirow}
\usepackage{psfrag}
\usepackage[linesnumbered,ruled]{algorithm2e}
\usepackage{soul} 
\usepackage[hyphens]{url}
\usepackage{hyperref}

\usepackage{float}
\usepackage{tikz}
\usetikzlibrary{arrows.meta, decorations.markings}
\usetikzlibrary{automata}
\usepackage{lipsum}  
\usepackage{subcaption}
\usepackage{xcolor}
\definecolor{boristext}{rgb}{0.3, 0.36, 0.88}
\definecolor{boriscomments}{rgb}{0.83, 0.0, 0.0}

\begin{document}


\begin{frontmatter}

\title{Cloud-gaming: Analysis of Google Stadia traffic}

\journal{Computer Communications}
\author{Marc Carrascosa \corref{cor1}}
\ead{marc.carrascosa@upf.edu}
\author{Boris Bellalta}
\ead{boris.bellalta@upf.edu}
\address{Wireless Networking Research Group, Universitat Pompeu Fabra \\ Carrer de Roc Boronat 138, 08018 Barcelona, Spain}
\cortext[cor1]{Corresponding author}

\begin{abstract}
Interactive, real-time, and high-quality cloud video games pose a serious challenge to the Internet due to simultaneous high-throughput and low round trip delay requirements. In this paper, we investigate the traffic characteristics of Stadia, the cloud-gaming solution from Google, which is likely to become one of the dominant players in the gaming sector. To do that, we design several experiments, and perform an extensive traffic measurement campaign to obtain all required data. Our first goal is to gather a deep understanding of Stadia traffic characteristics by identifying the different protocols involved for both signalling and video/audio contents, the traffic generation patterns, and the packet size and inter-packet time probability distributions. Then, our second goal is to understand how different Stadia games and configurations, such as the video codec and the video resolution selected, impact on the characteristics of the generated traffic. We also evaluate the ability of Stadia to adapt to different link capacity conditions, including cases where the capacity drops suddenly, as well as sudden increases in the network latency. Our results and findings, besides illustrating the characteristics of Stadia traffic, are also valuable for planning and dimensioning future networks, as well as for designing new resource management strategies. Finally, we compare Stadia traffic to other video streaming applications, showcasing the main differences between them, and introduce a traffic model using our captures. We show that this model can be used in simulations to further investigate the network performance in presence of Stadia traffic. 
\end{abstract}

\end{frontmatter}




\section{Introduction}

Video streaming is more popular than ever. It represents a share of 60.6\% of all internet downlink traffic, far above the second and third places: web browsing, which takes 13.1\%, and gaming  with 8.0\% \cite{Sandvine}. Video on demand services like Netflix, Amazon Prime Video and Disney+ report 183 million, 150 million and 50 million subscribers each \cite{forbes}, Youtube has 2 billions of logged-in users each month \cite{yt}, and Twitch reports an average of over 1.9 million concurrent users \cite{twitch}. 

Cloud gaming is the next step in streaming video content on demand, with several companies already offering subscription services that allow users to play video games remotely. In cloud gaming, the games are run in a remote server, and then streamed directly to the users, thus removing the need for dedicated gaming computers or consoles. While this has many advantages, it challenges the network infrastructure to support both high-throughput and low-latency, as otherwise the service will simply not run.

Some of the first companies to attempt this model were OnLive and Gaikai, which were later bought by Sony, who entered the market with PlayStation Now \cite{psnow}. Microsoft has its own service called XCloud, offering to play their games on Android phones and tablets \cite{xcloud}. Nvidia offers GeForce Now, allowing users to remotely play the games that they have bought previously in online stores on Windows, Mac, and Android devices \cite{geforce}. Lastly, Google has also entered the market with Stadia, a service that runs on Google Chrome or on Chromecast. Stadia has its own shop to purchase games and a subscription service offering free games each month. A key characteristic of Stadia is that it supports a 4K video resolution only for its PRO subscribers.

To deliver this service, Google uses Web Real Time Communication (WebRTC), an open standard by the World Wide Web Consortium (W3C), and the Internet Engineering Task Force (IETF), allowing peer-to-peer voice, video and data communication through browsers with a JavaScript API. WebRTC has a strong presence in videocalling and messaging. It is used by applications such as Google Hangouts or Amazon Chime, and it has the third spot in the global messaging market share after Skype and Whatsapp \cite{Sandvine}.


In this paper, we offer a comprehensible overview of how Stadia works, and the main characteristics of the traffic that Stadia generates. To do that, we perform a measurement campaign, obtaining a large dataset covering all aspects of interest, which also serves as a snapshot of Stadia behavior as a home user would perceive it near its launch\footnote{The dataset with all the collected traffic traces is available in open access here: \url{https://github.com/wn-upf/Stadia_cloud_gaming_dataset_2020}}. Our main goal is to characterize the traffic generated by Stadia, how it changes for different games, video codecs and video resolutions, and how Stadia reacts to different network conditions. To the best of our knowledge, this is the first work providing a detailed analysis of Stadia traffic, and likely, the first work analyzing the traffic generated by a cloud-gaming solution in such a detail.

The main findings of this paper are:
\begin{enumerate}
    \item Different games have different traffic characteristics such as the packet size, inter-packet times, and load. However, we have found the traffic generation process follows a common temporal pattern, which opens the door to develop general but parameterizable Stadia traffic models. We also found that some of these patterns are present in other WebRTC applications.
    \item The use of different video codecs (VP9 and H.264) and different video resolutions does not change the traffic generation process. Although we expected that the recent VP9 codec to significantly outperform H.264 in terms of traffic load, our results show they perform similarly, with H.264 even resulting in less generated traffic in some cases.  
    \item Stadia  works properly for different link capacities. We show that Stadia strives to keep the 1080p resolution at 60 frames per second (fps) even if the available bandwidth is far below its own pre-defined requirements, and only switches to a lower resolution of 720p as the last resort. In all cases, Round Trip Time (RTT) values  consistently remain below 25 ms, with average values between 10 and 15 ms, even when the experiments and measurements have been done at different times in a temporal span of several months.
    \item Stadia attempts to recover from a sudden drop in the available link capacity almost immediately, entering a transient phase  in  which Stadia aims to find a new configuration to compensate the lack of network resources. This transient phase however, can last over 200 seconds. During this time, although the user is able to continue playing, the quality of experience is heavily affected, with constant resolution changes, inconsistent framerate  and  even  audio  stuttering.  
    \item Stadia traffic adapts to sudden increases in network latency by increasing uplink traffic (i.e., client to server feedback), and decreasing the downlink video traffic. In contrast to the drops in bandwidth, the user experience is less affected when the latency increases, as no packets are dropped by the receiver and framerate is kept consistent for the whole capture.
    \item Traffic patterns found on Stadia are similar to those of other WebRTC applications, especially for video packets. For non-video packets, all the WebRTC applications tested show application specific patterns.

\end{enumerate}
 
Further, a traffic model is presented, which uses the patterns found in our captures such as the time between frames, packet size and number of packets that arrive in batches to accurately replicate Stadia traffic for a specific game (Tomb Raider) using different video resolutions. The model is built in a way it can be easily implemented as a traffic generator, either in a simulator or in the real world, which makes it suitable to be used in the performance evaluation of networking systems. Moreover, its parameters can be easily updated to cover a broad range of traffic generation patterns even if they do not belong to any particular game.

 The rest of the paper is organized as follows: Section \ref{sec:relwork} provides an overview of the related work. Section \ref{sec:webrtc} introduces Google Stadia, WebRTC and all protocols involved in their operation. The details of the experimental setup and definition of the datasets is presented in Section \ref{sec:exp}. A study of the bandwidth required to play multiple Google Stadia games can be found in Section \ref{sec:und_Stadia_trafficoverview}. An analysis of the main characteristics of Google Stadia traffic can be found in Section \ref{sec:und_Stadia_traffic}. Section \ref{sec:inside} studies the traffic evolution as the game changes states, and the effects of the video encoding and resolution can be found in Section~\ref{sec:video}. Performance under different available bandwidths is presented in Section~\ref{sec:band}, and the effects of such bandwidth changes on latency is investigated in Section~\ref{sec:latency}. Section~\ref{sec:latencychanges} analyses how Stadia adapts when latency increases suddenly. Section~\ref{sec:otherWebRTC} compares Stadia traffic patterns to those of other WebRTC applications and analyzes their similarities, and a model based on the traces is presented in Section~\ref{sec:ModelTR}. Finally, Section~\ref{sec:conclusions} closes the paper.

%


\section{Related Work}\label{sec:relwork}

Cloud gaming has received some attention, especially in regards to finding methods of compensating latency issues. The authors in \cite{latencyCloudGaming} present a crowdsourced study of 194 participants in which gameplay adaptation techniques are used to compensate up to 200 ms of delay. Some of these modifications include increasing the size of targets or reduce their movement speed, and the results show an improvement of the user QoE. If taken too far however, these latency adaptations reduce the perceived challenge of the game, resulting in an unsatisfying experience for users. An overview of the main issues with cloud gaming can be found in \cite{shea2013cloud}, where the performance of OnLive is tested. OnLive is one of the first cloud gaming services, highlighting the cloud overhead of 100~ms as a major challenge for player interaction with certain games. Authors in \cite{subjecti} emulate OnLive's system, streaming content from a PlayStation 3 and applying packet loss and delays to then perform a subjective study on the quality of the service perceived by users. As it could be expected, they observe that reduced delays are very important for fast paced games, while slower games suffer more from packet loss and degraded image quality than from latency issues. 
Lastly, the relationship between framerate and bitrate is studied in \cite{modelQoE}, presenting a QoE prediction model using both variables, and commenting that the  graphical complexity of different games is a challenge in generalizing such a model. Their results also show that for low bitrates, reducing the framerate leads to a better experience.

WebRTC performance on videoconferencing has been studied at length. The authors in \cite{singh2013performance} test the congestion control capabilities of WebRTC. Performing tests with different latencies, it is shown that performance is maintained while latency is below 200 ms. They also conclude that in presence of TCP cross-traffic, WebRTC video streams can heavily reduce their datarate to avoid an increase in latency at the cost of a lower video quality. The work in~\cite{webrtcstudy} does another in-depth analysis of the adaptability of WebRTC to different network conditions. WebRTC performance is evaluated in both wired and wireless networks, showing that the bursty losses and packet retransmissions from wireless connections have a severe impact on the video quality. WebRTC performance in presence of TCP downloads is studied in \cite{janczukowicz2016evaluation}, where multiple queue management methods are applied to avoid WebRTC performance degradation.

The quality of WebRTC services is assessed by defining two groups of metrics in \cite{garcia2019understanding}:
Quality of Service (QoS) ones such as latency, jitter, and throughput; and Quality of Experience (QoE) ones, which focus on the user satisfaction with the service. The evaluation of the QoE includes the use of both subjective methods, such as collecting feedback from users, and objective methods such as the Peak-to-Noise Ratio (PSNR) and the Structure Similarity (SSMI) index to calculate image degradation after compression and transmission. The work in \cite{ammar2016video} studies the use of Chrome's WebRTC internals as a source of QoS statistics, showing that the reported values for throughput and packet loss correlate well with user perceived issues in the connection. 

This paper aims to characterize the performance of Google Stadia, studying its traffic generation patterns under several different configurations, and analyzing its mechanisms for traffic adaptation. Similar studies of audio and video applications can be found in the literature. The work in \cite{bonfiglio2008detailed} studies how Skype changes its frame size, inter-packet gap and bandwidth according to network limitations, codec used and packet loss, noting on how packet size increases with them, retransmitting past blocks to compensate. The authors in \cite{suznjevic2014towards} used their own hardware to create a cloud gaming setup and study 18 different games, separated by genre, and find how said genre affects bandwidth, packet rate and video characteristics. In \cite{chen2013quality}, a study on cloud gaming platforms OnLive and StreamMyGame performed controlled experiments modifying the network delay, packet loss rate and bandwidth at the router to test and compare the performance of each service.

To be the best of our knowledge, this was the first paper focusing on the analysis of Stadia traffic when it was uploaded to arxiv in 2020 \cite{carrascosa2020cloudgaminganalysis}. Since then, it has served to other authors \cite{network1030015, 9484481, 9615562, 10.1145/3491043, 10.1145/3458335.3460963} as the starting point for their works on Stadia and other cloud gaming platforms. These papers provide complementary results to the analysis presented here by considering different network conditions, as well as particular network technologies such as Wi-Fi and mobile networks.


\section{How Stadia Works: WebRTC}\label{sec:webrtc}

In this Section we introduce how Stadia works. We first detail the user-server interaction, to then introduce the different protocols and mechanisms involved, such as the Google Congestion Control algorithm. Finally, we overview how ``negative latency" could have been implemented in Stadia, since to the best of our knowledge, there is no information available on this aspect.


\subsection{User-server interaction}

On a computer, Google Stadia can be played through Google Chrome. Once users reach \url{http://stadia.google.com}, they can either access the shop to acquire games (either by buying them or by just acquiring the ones provided for free with a PRO account), or go to their main page to choose one of the already available games to play. This part of Stadia is just regular web browsing until the user chooses a game, at which point the browser starts a WebRTC video session, switches to a full screen mode, and the user can start playing. 

Once the video session begins, the server transmits both video and audio, while the user transmits their inputs (coming from a gamepad or their mouse and keyboard). This way, both the video stream and the input stream have different traffic loads for each game: an action game will require constant inputs from the player, while a  puzzle game will have a slower and more methodical playstyle. Inside a game, since there are different states (menus, playing the game, idle, pause screen, etc.), traffic loads are also variable, although it is expected users will stay in the ``play" state for most of the time.

The stream of the game is customized according to the user's needs, and there are two parameters that have a direct impact in the quality of a stream: resolution and video codec. Stadia offers three different resolutions: 1280x720 (i.e., 720p), 1920x1080 (i.e., 1080p) and 3840x2160 (i.e., 2160p or 4K). The resolution can change mid stream automatically, according to the network state, but it can also be restricted by the users. Since higher resolution will require higher downlink traffic, Stadia allows users to restrict it to ``limited" (720p), ``balanced" (up to 1080p) and ``Optimal" (up to 4K). Note that Stadia only restricts the maximum resolution (i.e., a balanced configuration may still use 720p if network conditions are unfavorable).

The video encoding is selected automatically at the beginning of the session, and kept unchanged until it finishes. Stadia uses two video coding formats:

\begin{itemize}
    \item \textbf{H.264:}   It is the most supported video format nowadays, with 92\% of video developers using it in 2018, followed by its successor H.265 with 42\% \cite{videoReport}.  While H.265 promises half the bitrate of H.264 at the same visual quality~\cite{x265}\cite{x265_2}, H.264 has been supported in phones, tablets and browsers for years. Thus, for Stadia and other content providers, it serves as a fallback to ensure that no user will have issues decoding their media.
    \item \textbf{VP9:} Developed by Google in 2013 as the successor of VP8. It is royalty-free, as opposed to H.264 and H.265, and has a comparable performance to H.265~\cite{h265Comp}\cite{h265VP9}. It is already used by Youtube, and Google reports that VP8 and VP9 make up 90\% of WebRTC communications through Google Chrome~\cite{chromium}.
    
\end{itemize}

Audio encoding is done through \textbf{Opus} \cite{opusA}, an open audio codec released in 2012 and standardized by the IETF. Designed with voice over IP and live distributed music performances in mind, it can scale audio from 6 kbps to 510 kbps. It is used widely by applications such as WhatsApp\footnote{WhatsApp \url{https://www.whatsapp.com/}} and Jitsi\footnote{Jitsi \url{https://meet.jit.si/}}. 


\subsection{WebRTC}

\begin{figure*}[ht]
\centering
\includegraphics[width = 0.8\textwidth]{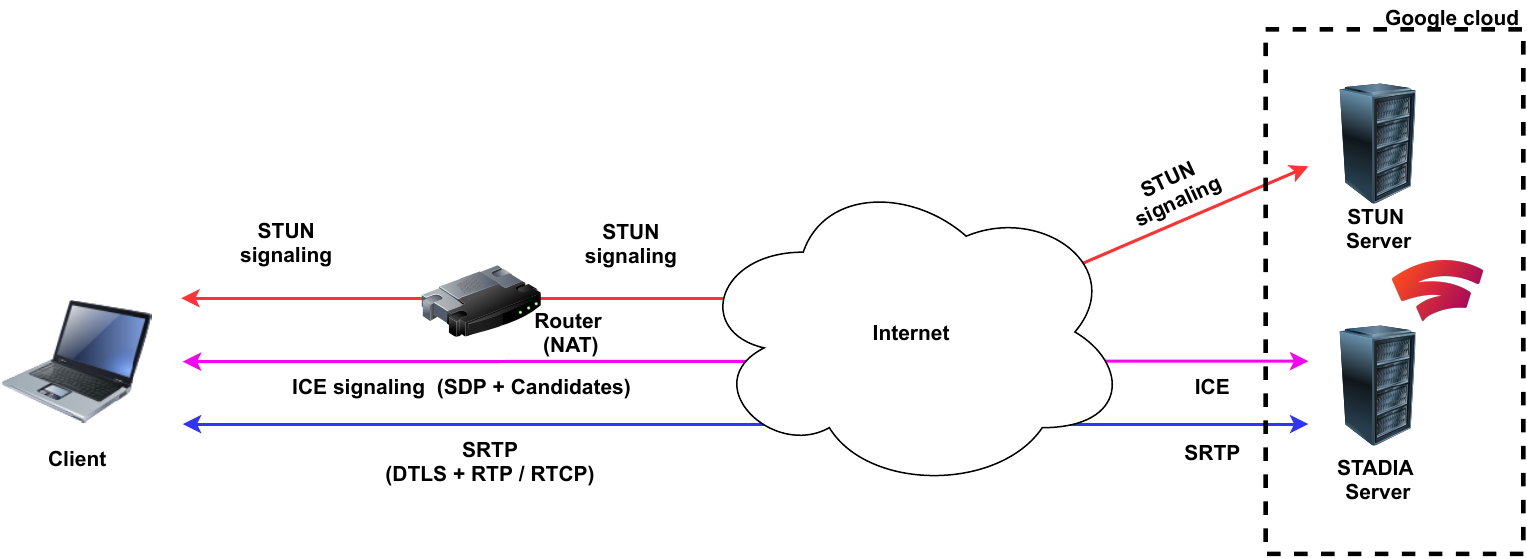}
\caption{Google's Stadia: Main components and data streams.}
\label{testbed}
\end{figure*}

Google Stadia uses WebRTC to provide its services, which uses  the following protocols:

\begin{enumerate}
    \item \textbf{ICE, STUN and TURN:} Interactive Connectivity Establishment (ICE) \cite{icerfc} is a protocol designed to facilitate peer-to-peer capabilities in UDP media sessions via Network Address Translator (NAT). It uses Session Traversal Utilities for NAT (STUN) and Traversal Using Relay NAT (TURN). STUN is used to query a server about the public IP of the user. Once both users know their respective IP and port, they can then share this information with the other user to establish a direct connection. TURN is used when a direct connection is not possible (because of a symmetric NAT). In such occasions, a TURN server is used as the intermediary between the users.
    
    In the signaling process, the Session Description Protocol (SDP) is used to negotiate the parameters for the session (audio and video codecs, IP and ports for the RTCP feedback, etc). These are exchanged along with the ICE candidates, which inform both peers of their connectivity options (IP and port, direct connection or through a TURN server, transport protocol used, etc).  UDP is the preferred protocol for most live streaming applications, but Transmission Control Protocol (TCP) is also supported by ICE.

    \item \textbf{DTLS:} Datagram Transport Layer Security (DTLS) \cite{dtlsrfc} \cite{srtp} is used to provide security in datagram based communications. It was designed as an implementation of TLS that did not require a reliable transport channel (i.e., it was designed for data exchanges that do not use TCP), as a consequence of the increased use of User Datagram Protocol (UDP) for live streaming applications, which prioritize timely reception over reliability. DTLS has become the standard for these kind of live streaming applications, and it is used in WebRTC to secure the RTP communication.
    \item \textbf{RTP and RTCP:} Real-Time Protocol (RTP) and Real-Time Control Protocol (RTCP)  \cite{rtprfc} \cite{rtpext} are used to transmit media over the secured channel. RTP is used by the sender of media (Stadia), while RTCP is used by the receiving client to provide feedback on the reception quality. RTP packets are usually sent through UDP, and contain a sequence number and a timestamp. The endpoints implement a jitter buffer, which is used to reorder packets, as well as to eliminate packet duplicates, or compensate for different reception timing.
    
    RTCP provides metrics that quantify the quality of the stream received. Some of these metrics are packet loss, jitter, latency, and the highest sequence number recieved in an RTP packet. RTCP packets are timed dynamically, and are recommended to account for only 5\% of RTP/RTCP traffic. 
    
\end{enumerate}

Once the ICE connection and DTLS encryption are in place, the connection consists mostly of RTP and RTCP packets for the video stream, and the application data (which we assume includes user inputs) sent through DTLS packets and STUN binding requests being sent periodically to ensure that the peer to peer connection can be maintained. RTP packets encrypted with DTLS maintain their structure, and do not use DTLS headers, which is why we can identify DTLS and RTP packets separately.


\subsection{Congestion Control}\label{gccex}

Google Congestion Control \cite{gcc} has two main elements: a delay-based controller at the client side, and a loss-based controller at the server side. The delay-based controller uses the delays between the video frame transmission at the sender and its arrival at the receiver to estimate the state of buffers along the path (i.e., it compares the time it took to transmit a full video frame at the sender, with the time it took for the sender to receive all packets that form that frame). Using this information, as well as the receiving bitrate in the last 500 ms, it calculates the required bitrate $A_r$, and forwards it to the sender. Notification messages from the client to the server are sent every second unless there is a significant change (i.e., a difference higher than 3\%) in the estimated bitrate  with respect to the previous one, in which case they are forwarded immediately. The loss-based controller works at the server side. It estimates the bitrate $A_s$ based on the fraction of packets lost provided by RTCP messages. Its operation is simple: If the packet loss is below 0.02, the $A_s$ is increased. On the contrary, if it is above 0.1, the bitrate $A_s$ is decreased. In case packet losses are in between 0.02 and 0.1, $A_s$ remains the same. The sender then uses the minimum of the two bitrates, i.e., $\min(A_r,A_s)$ to choose the current bitrate at which packets are transmitted.


\subsection{Negative Latency}

The mechanisms used by Stadia to manage network delays have been named \textit{negative latency} by Google. There is much speculation on what the term means, but the details have not been made available to the public. According to Google Stadia's designers: ``\textit{We created what we call a negative latency. It allows us to create a buffer for the game to be able to react with the user, and that accommodates the latency that you have on the network}" \cite{gNegL}. Their hardware infrastructure has also been mentioned \cite{Stadiainfra}, citing that 7500 Stadia edge nodes have been deployed at partnered ISPs to reduce the physical distance from server to user.

In \cite{outatime}, several aspects are considered to reduce latency in cloud gaming. Their approach consists of several aspects: a) Future state prediction with a Markov chain on inputs that they deem predictable, combined with error compensation for mispredictions (graphical rendering to quickly hide the misprediction and correct course); and b) Rollback: when a new input appears that contradicts a prediction, in-between frames are dropped to avoid propagating the error, this serves as a state recovery system, syncing the user and server when they drift. Basically, the system predicts future inputs from the user several frames before the input is taken. Then, these predictions (i.e., the video frame associated with each prediction) are transmitted to the client giving the illusion of instantaneous reaction. There are two types of inputs: ``predictive" and ``speculative". For ``predictive" inputs (user's movement) a Markov Chain is used to reduce the number of speculative frames to transmit.  For impulsive inputs they render multiple possibilities and send them all, only showing the correct one at the user side. Supersampling is also used (i.e., sampling the inputs of the user at a higher rate than the screen framerate), thus being able to receive multiple inputs per frame instead of limiting their inputs to one per frame. This reduces the sampling noise, and so improves the prediction accuracy. 

The authors in \cite{latencyCloudGaming} adjust the gameplay to compensate for latency issues on cloud gaming. They modify the size of the targets, their hitbox (i.e., the target is visually the same size, but the game registers hits on a larger area), their speed and quantity. These changes result in a better experience for most users. Authors in \cite{tanks} do the same, modifying a game of tank combat, and showing that latency severely affects those actions that require speed and precision.

Overall the consensus seems to be that ``negative latency" is a combination of closeness to the datacenter, predictive inputs, server side running at a higher frame rate for faster reactions, and parallel sending of speculative frames (i.e., many possible actions at once). 

\begin{figure*}[ht]
\centering
\begin{subfigure}[b]{0.28\textwidth}
\includegraphics[width = \textwidth]{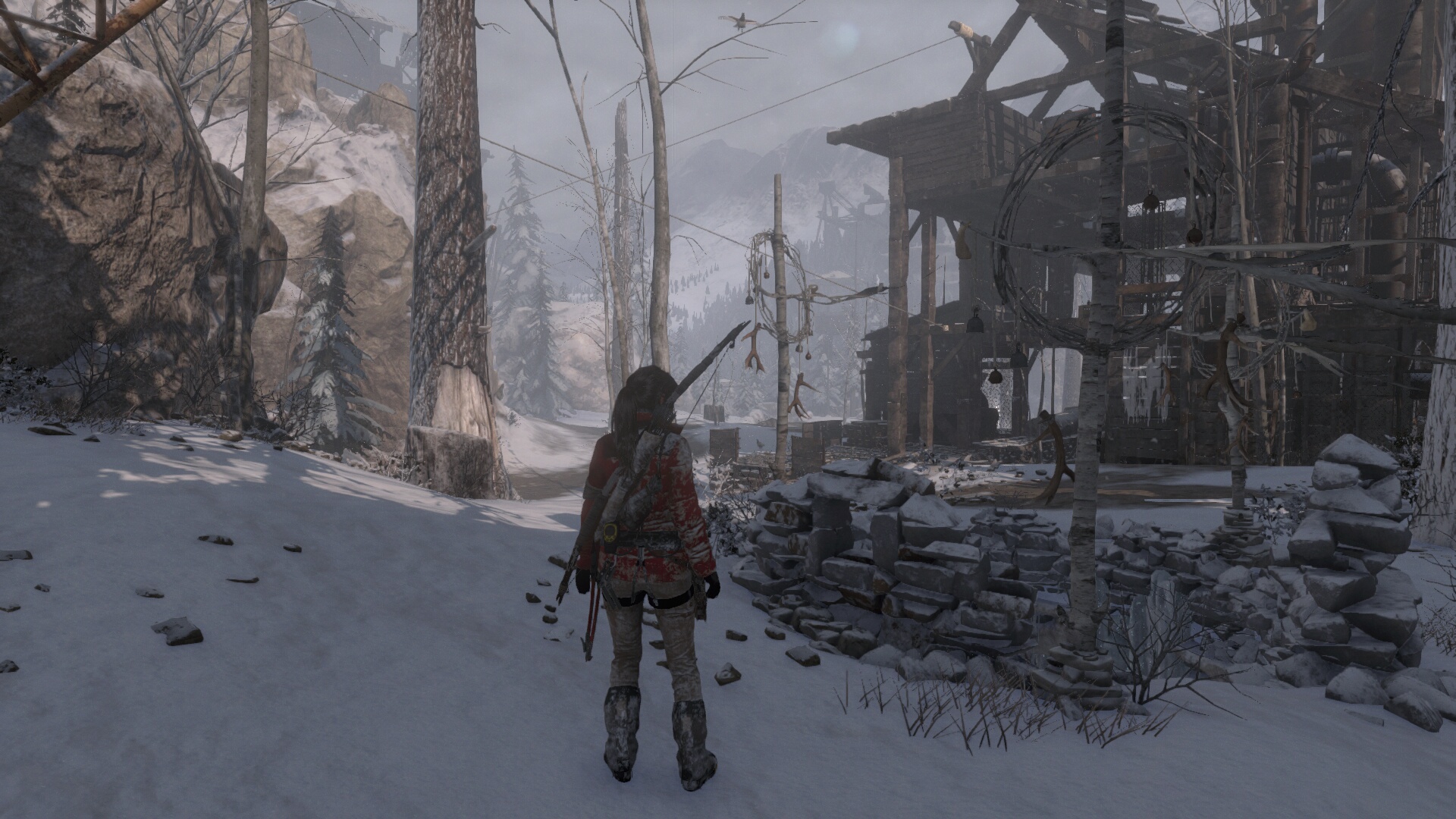}
\caption{Rise of the Tomb Raider}
\label{trsc}
\end{subfigure}
\begin{subfigure}[b]{0.28\textwidth}
\includegraphics[width = \textwidth]{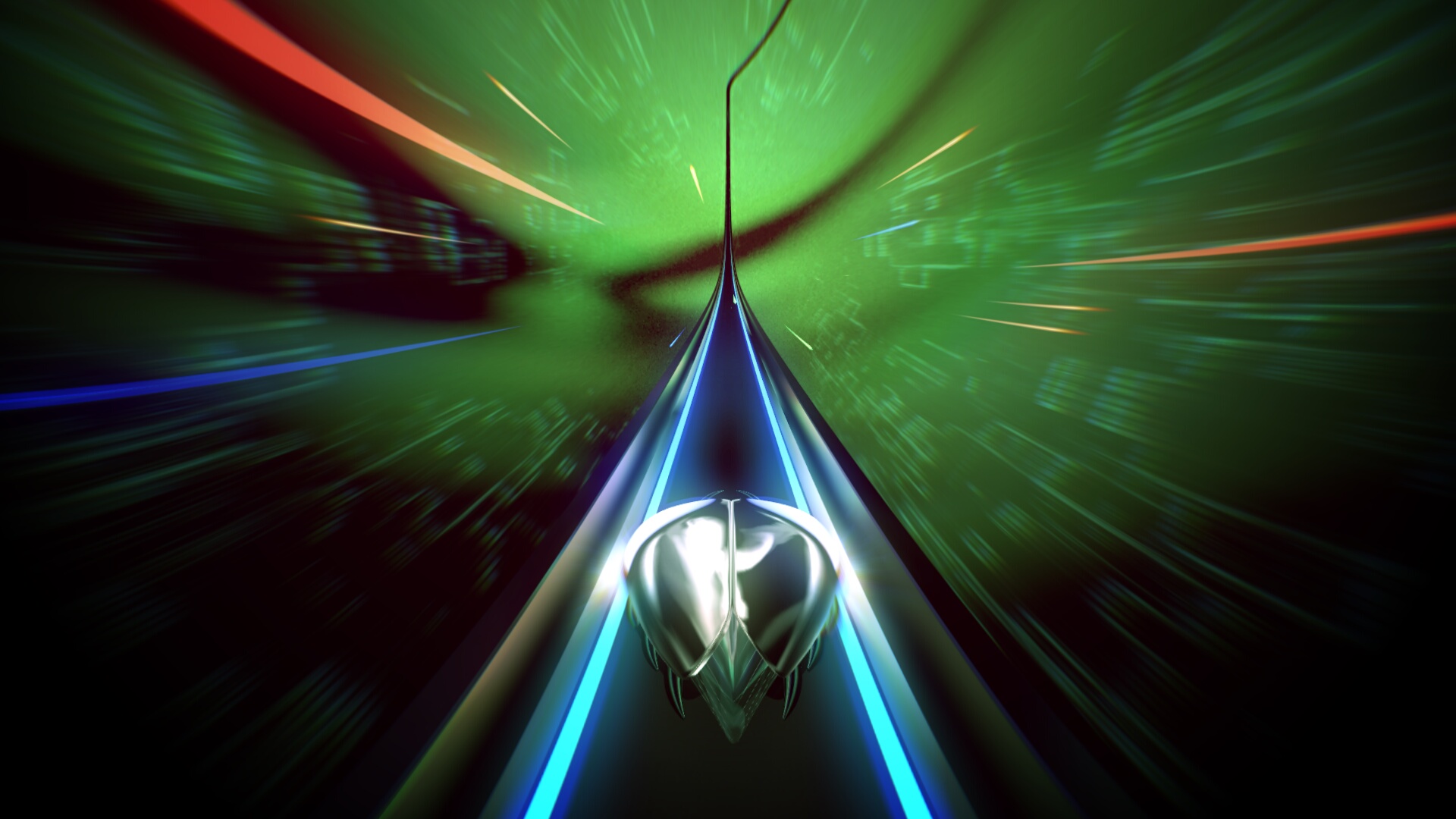}
\caption{Thumper}
\label{thsc}
\end{subfigure}
\begin{subfigure}[b]{0.28\textwidth}
\includegraphics[width = \textwidth]{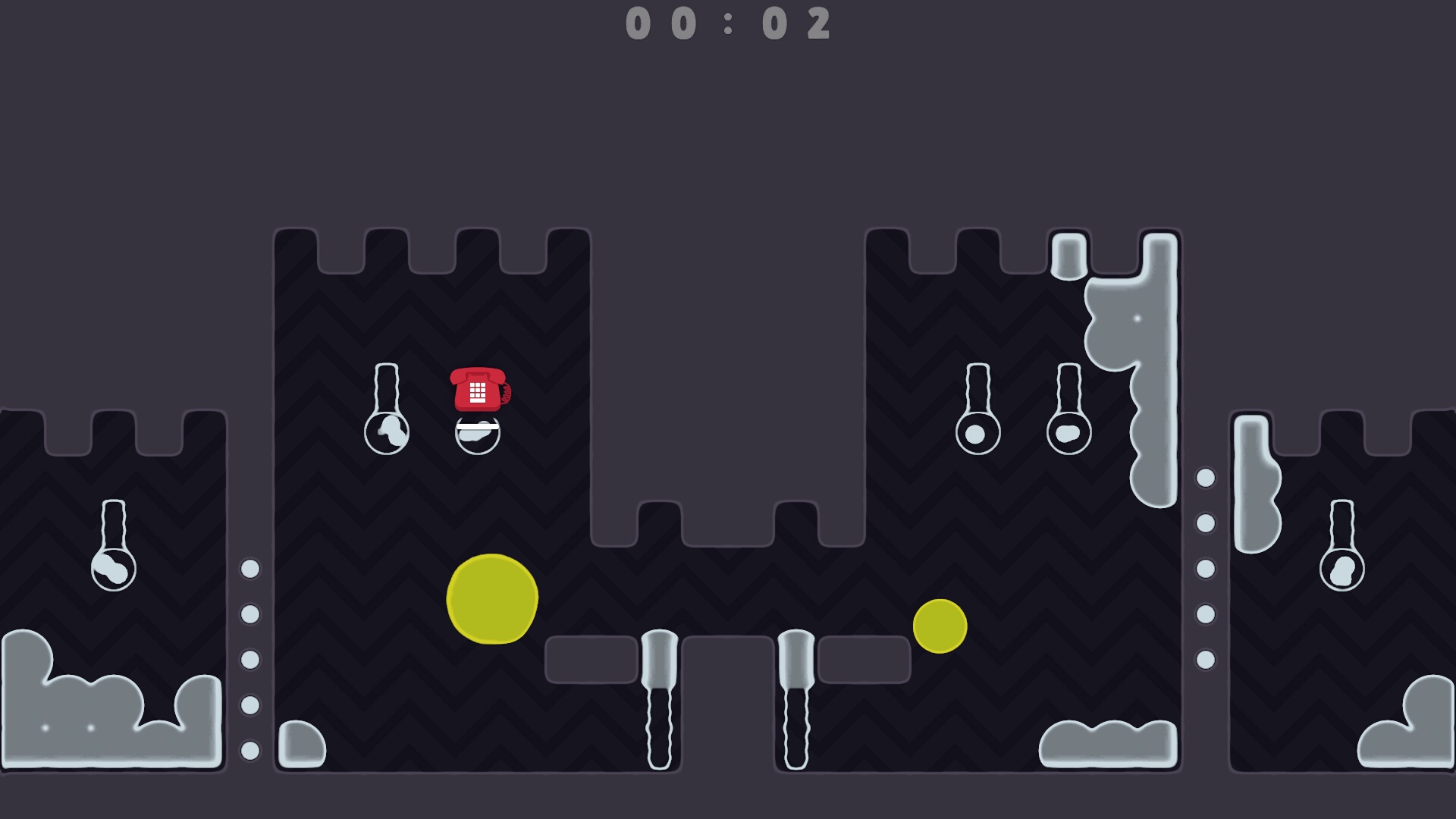}
\caption{Spitlings}
\label{spsc}
\end{subfigure}
\caption{Screenshots of the games considered in this paper.}
\label{Fig:Screenshots}

\end{figure*}


\section{Experiments and Measurements}\label{sec:exp}


This section introduces the considered testbed, the designed experiments, and the obtained datasets.


\subsection{Experimental Setup}

With the aim to identify the most relevant characteristics of Stadia traffic, we have designed a set of experiments to provide answers to the following five questions:
\begin{enumerate}
    \item How does Stadia generate the traffic? 
    \item Is the traffic constant regardless of the different game states?
    \item How does the selection of the video codec and resolution affect the generated traffic?
    \item Is Stadia able to adapt to different link/network capacities?
    \item How does Stadia react to a sudden network capacity change?
\end{enumerate}

For each experiment, we performed extensive traffic measurements while playing Stadia. All experiments were done in an apartment building of Sarria-St.Gervasi neighborhood, in the city of Barcelona, Catalonia, Spain.

We consider a network deployment as the one depicted in Figure~\ref{testbed}. The network consists of a laptop connected to a Wi-Fi-AP acting as an Ethernet switch.\footnote{To avoid any influence of the Wi-Fi channel on the measured traffic characteristics, we opted to do all tests using a CAT5 Ethernet cable.} The AP is a TP-Link Archer C7 v.5.0 running OpenWRT 19.07.2 \footnote{OpenWRT: \url{https://openwrt.org/}}. The client is a Dell Latitude 5580 running Ubuntu 16.04 (kernel 4.15.0-96) \footnote{Captures in 4K are run in Windows 10 due to Ubuntu drivers are not able to reach 60 fps.}. 

\setlength\tabcolsep{2 pt}
\renewcommand{\arraystretch}{1.1}

	\begin{table*}[ht]\centering
	\begin{small}
    \begin{tabular}{|c|c|c|l|}
  		\hline  
		\textbf{Dataset}  &\textbf{Name}&\textbf{Variables} &\multicolumn{1}{|c|}{\textbf{Characteristics}}  \\ \hline
        D1 & Temporal patterns & $Y_1, Y_2,Y_3 $ & TR, TH, SP; R: 1080p; C: VP9; T: 30 sec; DL \& UL; RTP, DTLS, STUN  \\\hline
        D2 & Traffic characteristics & $Y_1, Y_2,Y_3$ & TR, TH, SP; R: 1080p, C: VP9, T: 600 sec, DL \& UL, RTP, DTLS, STUN  \\\hline
        D3 & Game states & $Y_1, Y_2,Y_3$ & TR, SP;  R: 1080p, C: VP9, T: 540 sec, DL \& UL \\\hline
        D4 & Codecs & $Y_1, Y_2,Y_3$ & TR, SP; R: 1080p, C: VP9 \& H.264, T: 600 sec, DL\\\hline
        D5 & Resolutions & $Y_1, Y_2,Y_3$ & TR, SP; R: 720p, 1080p, 4K, C: VP9, T: 600 sec, DL   \\\hline
        D6 & Different bandwidths & $Y_1, Y_2,Y_3$ & TR, SP; R: 720p, 1080p, C: VP9, T: 60 sec, DL  \\\hline
        D7 & Sudden bandwidth changes & $Y_4,Y_5,Y_6,Y_7$  & TR, SP; R: 720p, 1080p, C: VP9, T: 500 sec, DL\\\hline
        D8 & Latency & $Y_7,Y_8$ & TR, SP; R: 720p, 1080p, C: VP9, L: 60-600 sec, DL \\\hline
        D9 & Latency changes & $Y_5, Y_6, Y_9, Y_{10}$ & TR: 1080p, C: VP9, L: 180 sec, DL \& UL \\\hline
	\end{tabular}	\caption{Variables included in each file and dataset. R: resolution, C: video codec, T: duration, DL: downlink traffic, and UL: uplink traffic. In case the trace includes only packets from a specific protocol, it is also indicated. }
	\label{ds}
	\end{small}
	\end{table*}

All tests are performed using the Chrome browser (version 81.0.4044.92), with Wireshark (version 3.2.2)\footnote{Wireshark: \url{https://www.wireshark.org/}} running in the background to capture all incoming and outgoing traffic. Wireshark captures are processed via tshark to extract a .txt file that is later interpreted via MATLAB 2019a \footnote{Matlab: \url{https://mathworks.com/products/matlab.html}}. Filters include all RTP, RTCP, DTLS and STUN frames, and we extract the arrival timestamps, inter-packet time and packet size. Once the capture is finished, WebRTC performance metrics (number of frames decoded, codecs used, resolution, round trip time, and jitter buffer) are extracted from Chrome via \url{chrome://webrtc-internals}.

Apart from Section \ref{sec:und_Stadia_trafficoverview} in which we compare the bandwidth requirements of multiple Stadia games, all other experiments are based on the following three games: Rise of the Tomb Raider: 20th anniversary edition (TR), Thumper (TH) and Spitlings (SP). Tomb Raider is a third person action adventure game with an open world and a lot of freedom on player inputs, as well high definition graphics that require high throughput. Thumper is an on rails rhythm game, with more limited player inputs and predictable gameplay, contrasting the freedom on TR while still requiring high throughput. Spitlings is a 2D platformer, and the game that required the lowest throughput out of all the games we tested. An illustrative screenshot of each game can be found in Figure~\ref{Fig:Screenshots}.

Although we perform the measurements at the client side, we assume the network between server and client is of high capacity, and therefore its effects on shaping the traffic are negligible, and do not significantly alter Stadia traffic characteristics. This assumption is later discussed in Section~\ref{sec:und_Stadia_traffic}, where, from the collected traffic traces, we conjecture it is accurate.


\subsection{Datasets} \label{sec:datasets}

After the experiments and measurement campaign, we have generated nine datasets\footnote{The datasets are publicly available at Github: \href{https://github.com/wn-upf/Stadia_cloud_gaming_dataset_2020}{https://github.com/wn-upf/Stadia\_cloud\_gaming\_dataset\_2020} }. Each dataset can contain multiple traffic traces depending on the number of experiments carried out in each category. Each traffic trace is a text file that includes two or more of the following variables as columns:

\begin{itemize}
    \item $Y_1$: Packet arrival time in Epoch format.
    \item $Y_2$: Arrival time relative to previous packet in seconds.  
    \item $Y_3$: Length of UDP payload in bytes.
    \item $Y_4$: Video frame height in pixels.   
    \item $Y_5$: Video frames per second.  
    \item $Y_6$: Round Trip Time in seconds. 
    \item $Y_7$: Packets lost per second.   
    \item $Y_8$: Jitter Buffer Delay per second.
    \item $Y_9$: Uplink RTCP data in bits per second.
    \item $Y_{10}$: Downlink RTP data in bits per second.
    
\end{itemize}

Variables $Y_1$, $Y_2$ and $Y_3$ are obtained via Wireshark, and correspond to the filters \textit{frame.time\textunderscore epoch}, \textit{frame.time\textunderscore delta\textunderscore displayed} and \textit{udp.length}. The rest of the variables are extracted via webRTC internals: receiver frame height and jitter buffer delay are taken from \textit{RTCMediaStreamTrack}, frames decoded per second, packets lost and downlink RTP data from \textit{RTCInboundRTPVideoStream}, uplink RTCP data from \textit{RTCDataChannel} and round trip time from \textit{RTCICECandidatePair}.

Table \ref{ds} specifies which of the previous variables are used in each of the datasets, as well as information of each dataset, such as the games, the duration of the measurement, the video codec used, and the resolution, among other aspects.


\section{Games Overview} \label{sec:und_Stadia_trafficoverview}

The following Sections, including this one, focus on the analysis of Stadia traffic. The main goal is to identify the main characteristics of Stadia traffic, and determine how it changes for different games, video codecs and video resolutions. Moreover, we will also focus on how Stadia adapts its traffic to different and changing network conditions. 

In this section we overview 10 games available through the Stadia pro subscription (their name and genre can be found in Table \ref{sutaula}). We perform captures at a video resolution of 1080p for all games, selecting only 60 seconds in which the game is being played for the analysis. Here, we compare the throughput of these 10 games for both uplink and downlink.
	\setlength\tabcolsep{4 pt}
\renewcommand{\arraystretch}{1.2} 

	\begin{table}[t]\centering
	\begin{small}
    \begin{tabular}{|c|c|c|c|c|c|c|}
  		\hline  
		\textbf{Game}&  \textbf{Genre  }\\\hline
		
	  Tomb Raider  & Open world action-adventure\\\hline
      Thumper  & On-rails ryhtm game  \\\hline
      Spitlings  & Side-scrolling platformer   \\\hline	
      Gylt  & Action-adventure   \\\hline
      Grid  & Racing   \\\hline
      SuperHot  & First person shooter   \\\hline
      Samurai Shodown & 1 on 1 fighting  \\\hline
      Serious Sam 3  & First person shooter  \\\hline
      Farming Simulator 2019  & Simulation    \\\hline
      Panzer Dragoon & On-rails third person shooter \\\hline
	\end{tabular}	
	\caption{Games considered and their respective genre.  }
	\label{sutaula} 
	\end{small}
	\end{table}

\begin{figure*}[t!]
\centering 
\begin{subfigure}[b]{0.49\textwidth}
\includegraphics[width=\textwidth]{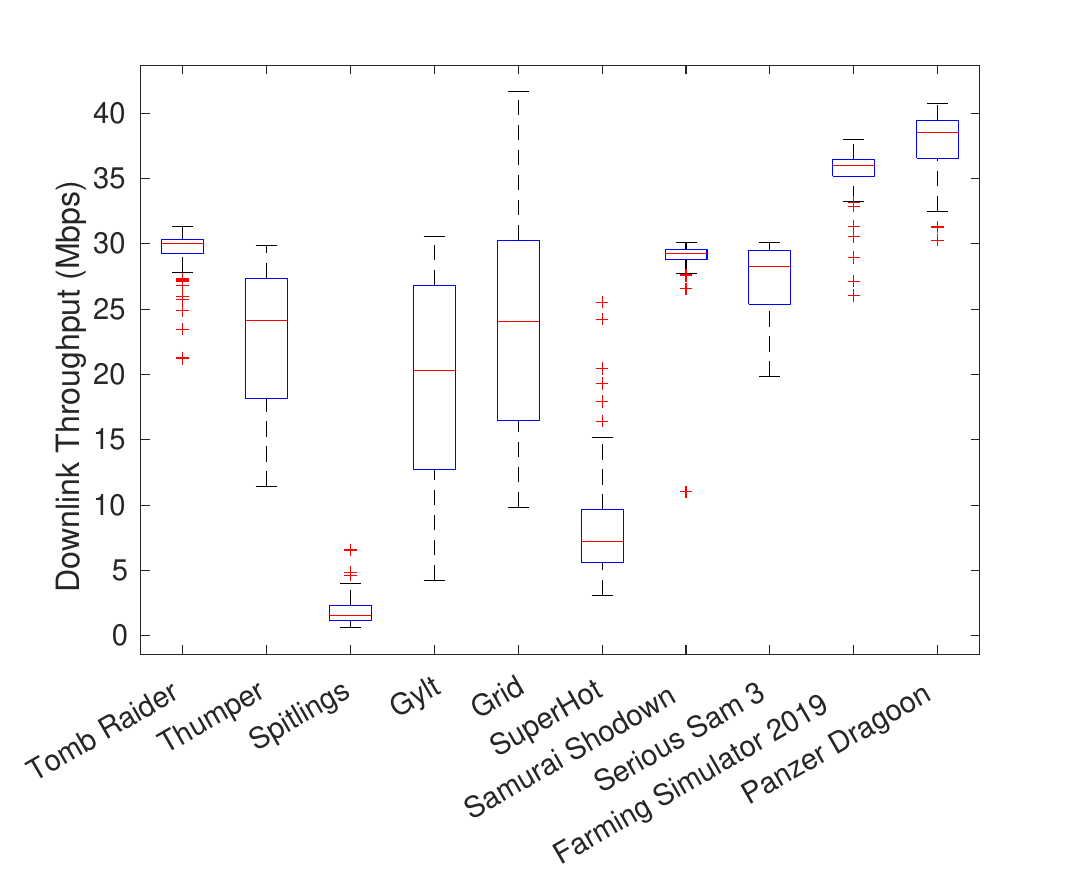}
\caption{Downlink traffic}
 \label{dlt}
\end{subfigure}
\begin{subfigure}[b]{0.49\textwidth}
\includegraphics[width=\textwidth]{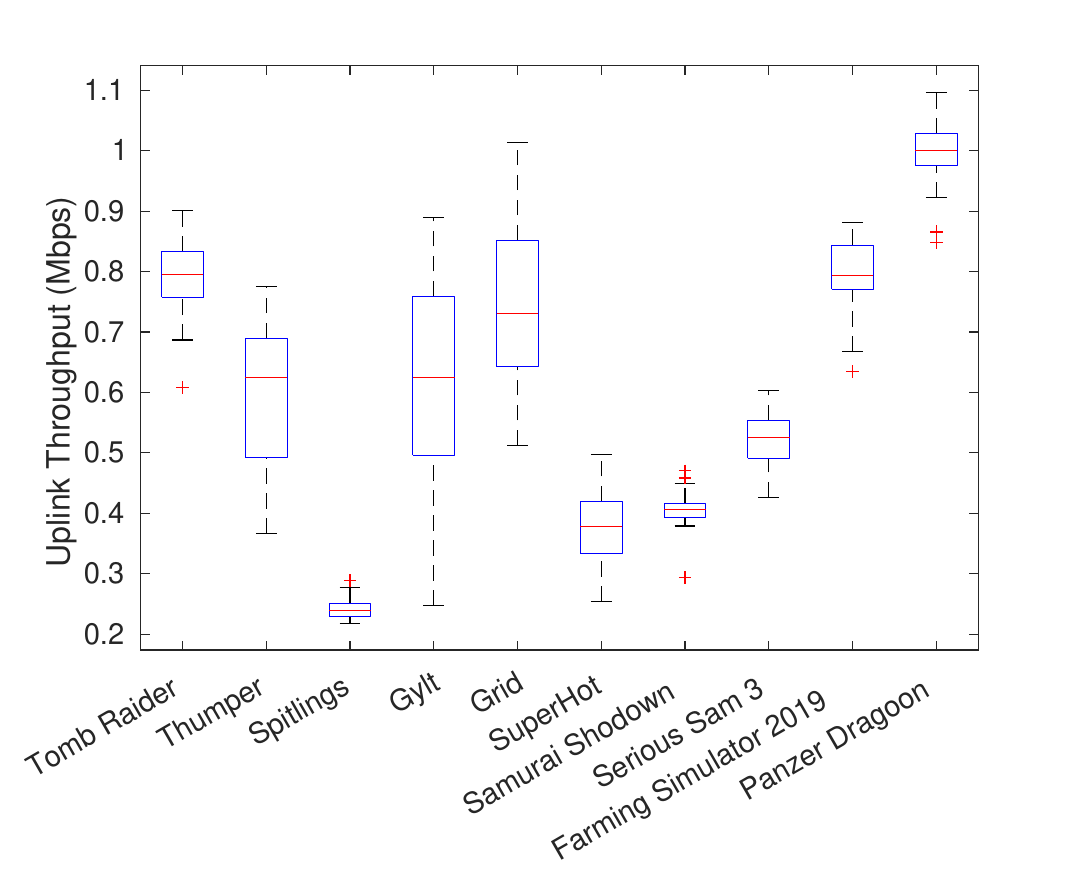}
    \caption{Uplink traffic}
    \label{ultstadia}
\end{subfigure}
\caption{Boxplot of the throughput for different games. }
\label{t}
\end{figure*}

Figure \ref{dlt} shows the boxplot for downlink UDP traffic in Mbps for each of the games tested. Each game requires a different amount of traffic, and some games have a much higher variance than others. It can be observed that most of the games require a heavy downlink load, as 80\% of them have the 25th percentile over 10 Mbps and their median over 20 Mbps. The highest variance can be found on the racing game Grid, as it shows values ranging from $9.83$ Mbps to $41.6$ Mbps (standard deviation of $8.6$), while the 2D platformer Spitlings shows the lowest, ranging only from $0.645$ Mbps to $6.56$ Mbps (standard deviation of $1.2$). Figure \ref{ultstadia} shows the uplink UDP traffic in Mbps. While the boxplots show similar shapes to the downlink ones, the traffic values are much lower, as all games stay below $1.1$~Mbps.

\textit{Finding:} Every Stadia game has different bandwidth requirements, with average traffic going from $1.95$ Mbps to $37.7$ Mbps depending on the game. In general, most games seem to demand a high amount of traffic, but there are some exceptions such as Spitlings. Uplink traffic is in average 37.63 times lower than downlink traffic, ranging from $0.21$ Mbps to $1.1$ Mbps depending on the game.

\setlength\tabcolsep{4 pt}
\renewcommand{\arraystretch}{1.2} 

	\begin{table}[ht!!!]\centering
	\begin{small}
    \begin{tabular}{|c|c|c|c|}
  		\hline  
		\multirow{2}{*}{\textbf{Parameter}}&  	\multirow{2}{*}{\textbf{Avg. Packet  }} & 	\multirow{2}{*}{\textbf{Avg. inter } }& 	\multirow{2}{*}{\textbf{Load }}\\[5pt] 
		 &\textbf{size (bytes)} & \textbf{packet  time (ms)}& \textbf{(Mbps)}\\\hline
		\multicolumn{4}{|c|}{\textbf{Downlink}}\\\hline
		
	  TR RTP  &  1118.01   & 0.34    & 25.60     \\\hline
      TH RTP  & 1154.64    & 0.49    & 18.33     \\\hline
      SP RTP  & 677.21   & 2.81    & 1.87     \\\hline	
      TR STUN  &  81.39   & 265.23    & 0.0024     \\\hline
      TH STUN  & 81.50    & 263.31    & 0.0024     \\\hline
      SP STUN  & 81.50    & 264.36    & 0.0024     \\\hline
      TR DTLS  &  118.59  & 7.44     & 0.12      \\\hline
      TH DTLS  &  132.44  & 10.52    & 0.097      \\\hline
      SP DTLS  &    137.38 & 11.31     &    0.094   \\\hline\hline
     
      \multicolumn{4}{|c|}{\textbf{Uplink}}\\\hline
      
      TR RTCP  &  65.99   & 1.44    & 0.35     \\\hline
      TH RTCP  & 65.99    & 1.98    & 0.26     \\\hline
      SP RTCP  & 113.76    & 9.84    & 0.090     \\\hline
      TR STUN  &  79.37 & 265.13 &  0.0024   \\\hline
      TH STUN  & 79.25 & 261.04 &  0.0023 \\\hline
      SP STUN  & 79.10 & 264.35  &  0.0023 \\\hline
      TR DTLS  & 123.17 & 7.10 & 0.13\\\hline
      TH DTLS  & 114.66 & 9.96 & 0.089  \\\hline
      SP DTLS  & 119.60 & 10.62 & 0.087 \\\hline
     
	\end{tabular}	
	\caption{Traffic characteristics for RTP/RTCP, DTLS and STUN streams. TR: Tomb Raider, TH: Thumper, SP: Spitlings. }
	\label{supertaula} 
	\end{small}
	\end{table}


\section{Stadia traffic} \label{sec:und_Stadia_traffic}

	\begin{figure*}[ht]
    \centering
    \includegraphics[width = \textwidth]{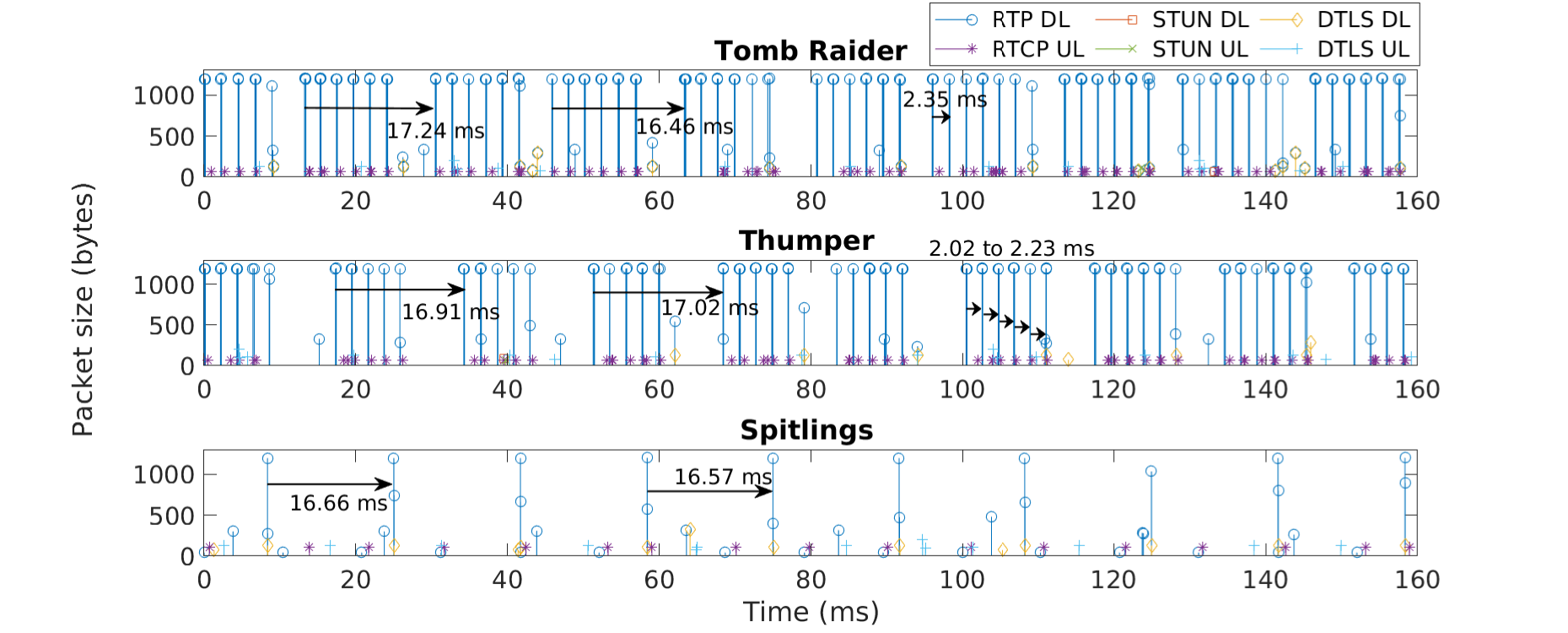}
    \caption{Temporal evolution of Stadia traffic for Tomb Raider, Thumber and Spitlings.}
    \label{separation2}
    \end{figure*}

\begin{figure*}[ht!!!]
\centering 
\begin{subfigure}[b]{0.329\textwidth}
\includegraphics[width=\textwidth]{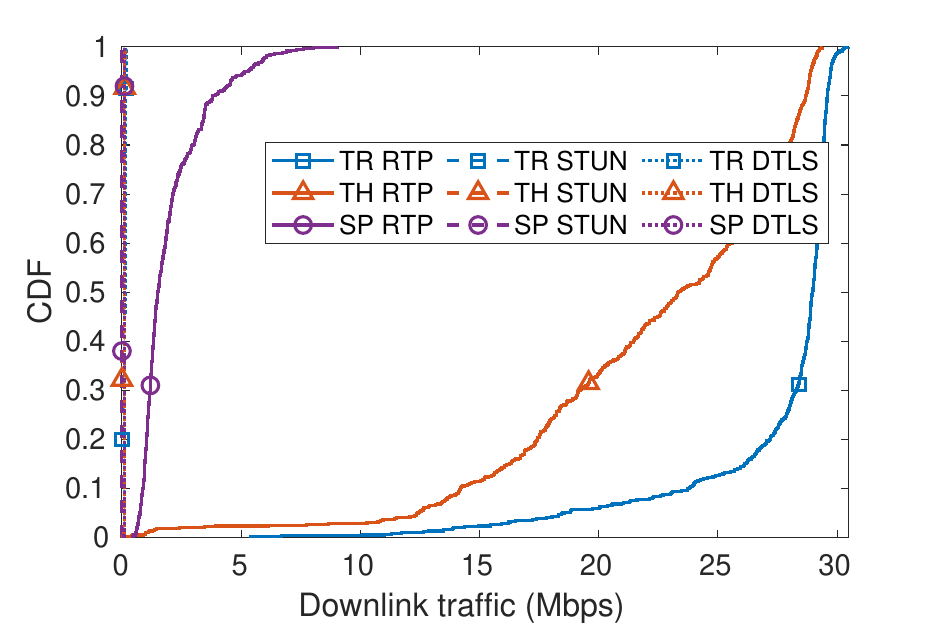}
\caption{Downlink traffic}
 \label{dlvul}
\end{subfigure}
\begin{subfigure}[b]{0.329\textwidth}
\includegraphics[width=\textwidth]{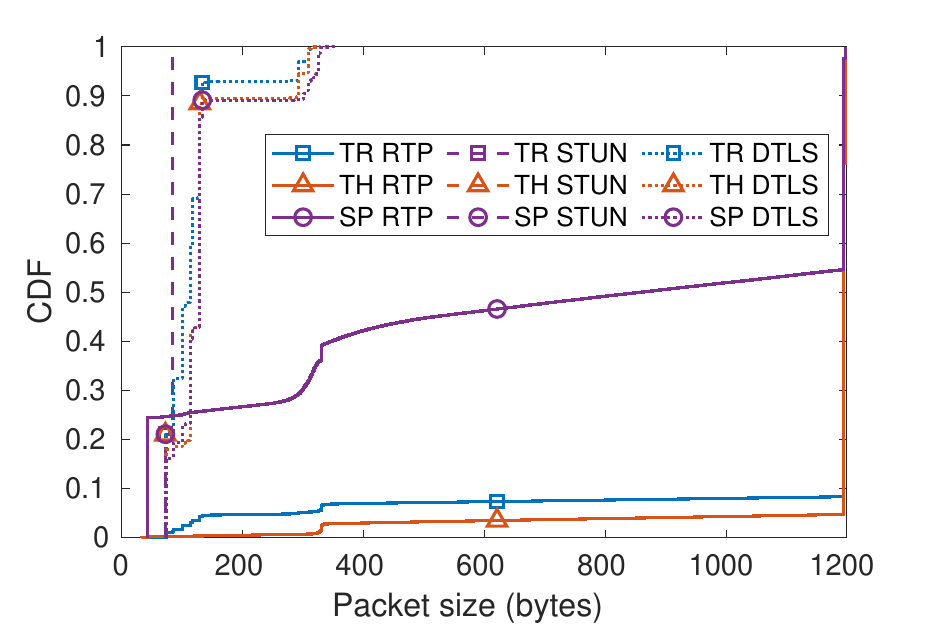}
    \caption{Packet size}
    \label{byprot}
\end{subfigure}
\begin{subfigure}[b]{0.329\textwidth}
\includegraphics[width=\textwidth]{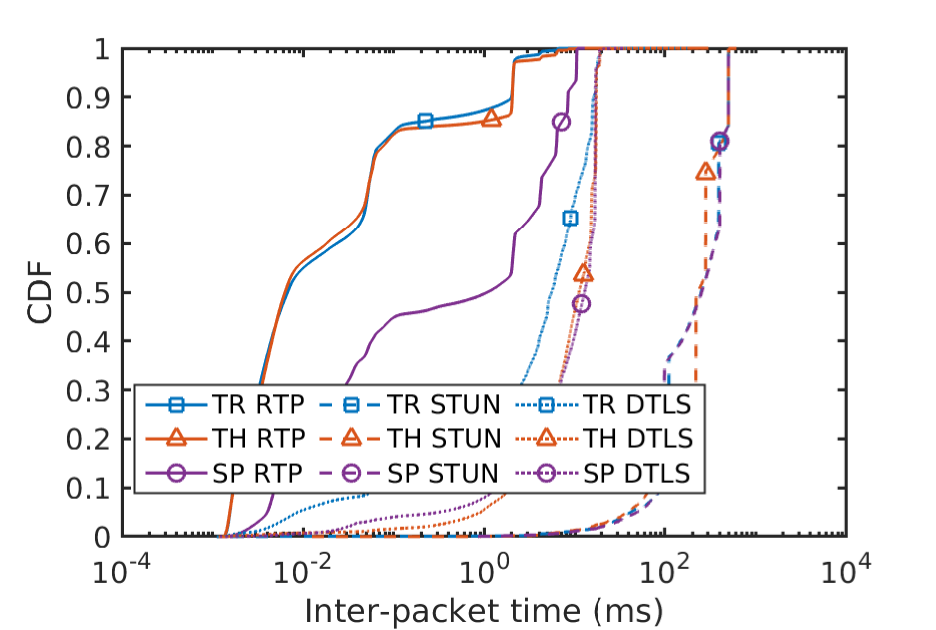}
    \caption{Inter-packet time}
    \label{ifcdf1}
\end{subfigure}
\caption{CDFs of RTP, DTLS and STUN in the downlink, for TR, TH and SP.}
\label{one}
\end{figure*}

In this Section, we first determine the traffic share between RTP, DTLS and STUN streams, showing that, as expected, in terms of network utilization, only RTP traffic is relevant. We also evidence the existence of temporal patterns in the generated traffic, as well as the existing correlation between downlink and uplink streams. We finally delve in the packet size, inter-packet time and traffic load characteristics. Starting with this Section we will focus on the three main games mentioned in Section \ref{sec:exp}: Tomb Raider, Thumper and Spitlings.


\subsection{RTP, DTLS and STUN streams}

We are interested in quantifying the fraction of traffic that belongs to RTP/RTCP, DTLS and STUN streams.  We use dataset D1 (Table \ref{ds} in Section \ref{sec:datasets}). It includes three different games: Tomb Raider, Thumper and Spitlings. While Tomb Raider and Thumper are 3D games, Spitlings is a 2D game. Therefore, we expect Tomb Raider and Thumper will require higher network resources than Spitlings. Note that D1 includes only traces of the games in the ``play" state.

Table \ref{supertaula} shows the average packet size, average inter-packet time, and average traffic load of RTP, RTCP, STUN and DTLS streams for the three considered games. It can be seen that most of the traffic corresponds to RTP as it carries the game video and audio contents from the server to the client. Note that while more than 90\% of all traffic corresponds to the downlink (91.64\% for Spitlings, and 98.13\% and 98.14\% for Tomb Raider and Thumper), the uplink can take up to 30.81\% of all transmitted packets, showing that Stadia requires a consistent stream of short periodic reports to function, even if their overall network utilization seems negligible.

\textit{Finding:} RTP/RTCP traffic is the only stream of Stadia traffic in which the packet size, inter-packet times, and traffic load depend on the game played. DTLS and STUN traffic are game-independent traffic streams.


\subsection{Temporal patterns}

We aim to visualize how Stadia traffic looks like over time. We use dataset D2. This time we focus on a small snapshot of the dataset to visualize the temporal traffic evolution.

Figure~\ref{separation2} shows the traffic evolution of Tomb Raider, Thumper and Spitlings during 160 ms. First, in all three games, and focusing only on the downlink RTP packets, we can observe a clear pattern that repeats every $\approx$16.67~ms, and that corresponds to the video frame rate of Stadia (i.e., 1/60 fps)\footnote{The same frame-based pattern can be observed in all other games tested.}. Second, between two consecutive frames, we can observe several groups of packets separated by 2 ms. The number of groups and the number of packets in each group depend on the game. For Tomb Raider, there are 6 groups of 7 to 9 packets each, while for Spitlings, there is only 1 group of 1-2 packets each. Thumper has 6 groups of 5 to 7 packets each. The existence of those groups may be due to the generation of large video frames at the source, which need to be spread among multiple packets. However, since values  change for different games, we conjecture this is already implemented at the source. 
The smaller RTP packets of around 360 bytes that appear with a periodicity of 20 ms, represent the audio stream, which has a bitrate  $\approx$120 kbps in all three games. We observe the same patterns through all the traffic captures beyond the 160~ms shown in Figure~\ref{separation2}, and this consistency allows us to surmise that they are not affected by the transport network. Lastly, the existing correlation between downlink RTP and uplink RTCP streams is clearly observable.

\textit{Finding:}  The temporal evolution of RTP/RTCP traffic follows a well-defined pattern. Inside each frame period, RTP packets are sent in groups. The number of groups and the size of each group in number of packets depend on the game. 

\subsection{Traffic load, packet sizes and inter-packet times} \label{sec:traffic_char}

We have seen that the temporal structure of Stadia traffic follows a clear periodic pattern. Here, we aim to further validate previous results by showing the probability distribution of the packet size, the inter-packet delay, and traffic load. 
We use the complete dataset D2 that covers 10 minutes of gameplay for each game.

Figure~\ref{dlvul} shows the ecdf of the traffic load for each game and protocol, where we can observe that both STUN and DTLS traffic represent a small fraction of the total traffic generated by Stadia, with a maximum of 3.2~Kbps and 159.9~Kbps, respectively. As we have seen before, RTP represents most of the traffic, with Tomb Raider and Thumper generating, respectively, 28.97 and 23.32 Mbps at the 50th percentile. Spitlings generates much lower traffic loads, with only 1.5 Mbps at the same percentile. Figure~\ref{byprot} shows the ecdf of the packet size. For Tomb Raider and Thumper, more than 90\% of the RTP packets are equal or larger than 1194 bytes, while for Spitlings these larger packets only represent the 45.42\%. For STUN and DTLS, we observe they transmit very small packets. The average packet size of STUN is 81.47 bytes for Tomb Raider, 81.48 bytes for Thumper and 81.47 bytes for Spitlings. For DTLS, their average packet size is 114.47, 133.36 and 129.24 bytes respectively.  Finally, Figure~\ref{ifcdf1} shows the inter-packet time for each game and protocol. As expected, inter-packet times are inversely proportional to the traffic load, with Tomb Raider and Thumper showing that, respectively, the 87.33\% and 85.07\% of their RTP packets have an inter-packet time below 1~ms. Spitlings has higher inter-packet times in general, with only 49.72\% of them below 1~ms. 

We can also observe that Spitlings seems to have consistent low traffic, with occasional peaks, while Tomb Raider is the opposite. It generates a high traffic load in general, with occasional dips.

\textit{Finding:} In this section, by comparing the different probability distributions, we further confirm that DTLS and STUN traffic are almost identical regardless of the game. With respect to the RTP/RTCP traffic, similarly, we also confirm that the video traffic generation process is common in all three games, just adapting for each game the number of groups of packets per frame, and the number of packets inside each group. Although this paper does not focus on traffic modelling, these results open the door to develop a general but parameterizable traffic model for Stadia traffic.


\section{Inside a game}\label{sec:inside}

In the previous section we showed that the traffic generated by Stadia in the ``play" state depends on the game. However, in addition to the ``play" state, a game has other states, such as the ``menu", ``idle", and ``pause". Here, we investigate how is the traffic generated in each state of a game.


\subsection{Different game states, different traffic loads}

In this section we check how the traffic load generated by Stadia changes based on the different states (or types of screens) that we can find in a game. We also study how user input changes the load perceived.

We use dataset D3, that includes traces from Tomb Raider and Spitlings. We omit Thumper, as we have seen before that it behaves similar to Tomb Raider. We play both games at 1080p using the VP9 codec, and change the game state every 120~seconds. We consider a state to be a ``screen" that shows unique characteristics in both gameplay and video needs. For example, the main menu is text based, with an animated background, and the player just selects items from a list. This is different than the game itself, in which there is a changing environment, with many active elements on the screen, and multiple possible actions for the player. We have identified the following states: 

\begin{enumerate}
    \item \textbf{Main Menu:} A screen with several text items (e.g., new game, continue, options, etc.), and a dynamic background (i.e., an office with lightning in the windows for Tomb Raider, and several characters moving around for Spitlings).
    \item \textbf{Loading screen:} A black screen with some text that appears while a level is loading.
    \item \textbf{Idle:} Inside of the game state, but with the player not performing any actions.
    \item\textbf{Play:} The player interacts with the environment by taking actions.
    \item \textbf{Pause:} The pause menu is another text menu, overlayed on top of the frozen playing screen.
\end{enumerate}

\begin{figure}[t!]
    \centering
    \includegraphics[width=0.4\textwidth]{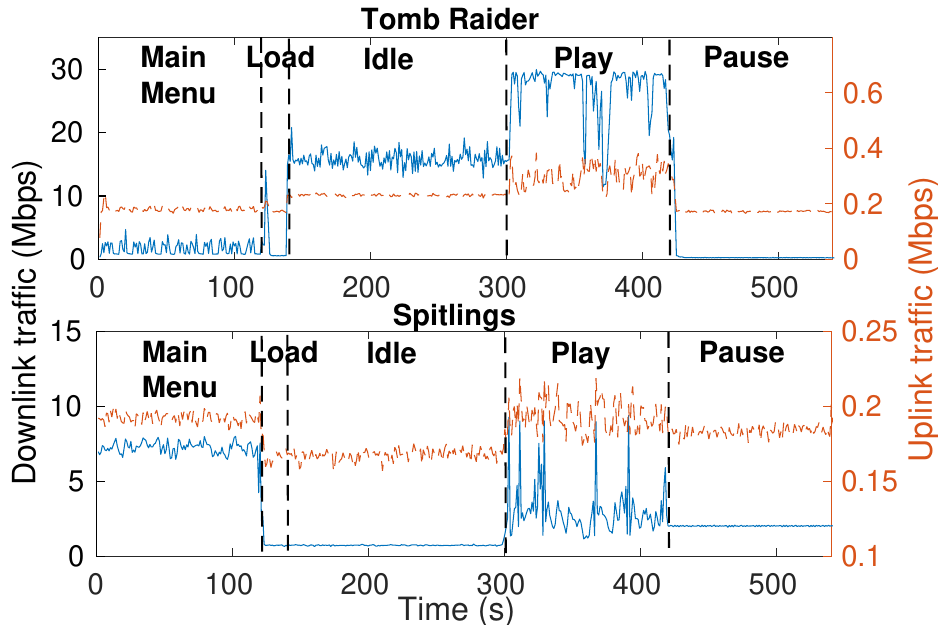}
    \caption{Traffic evolution over 4 states. }
    \label{states}
\end{figure}

\begin{figure*}[ht!!]
\centering 
\begin{subfigure}[b]{0.329\textwidth}
\includegraphics[width=\textwidth]{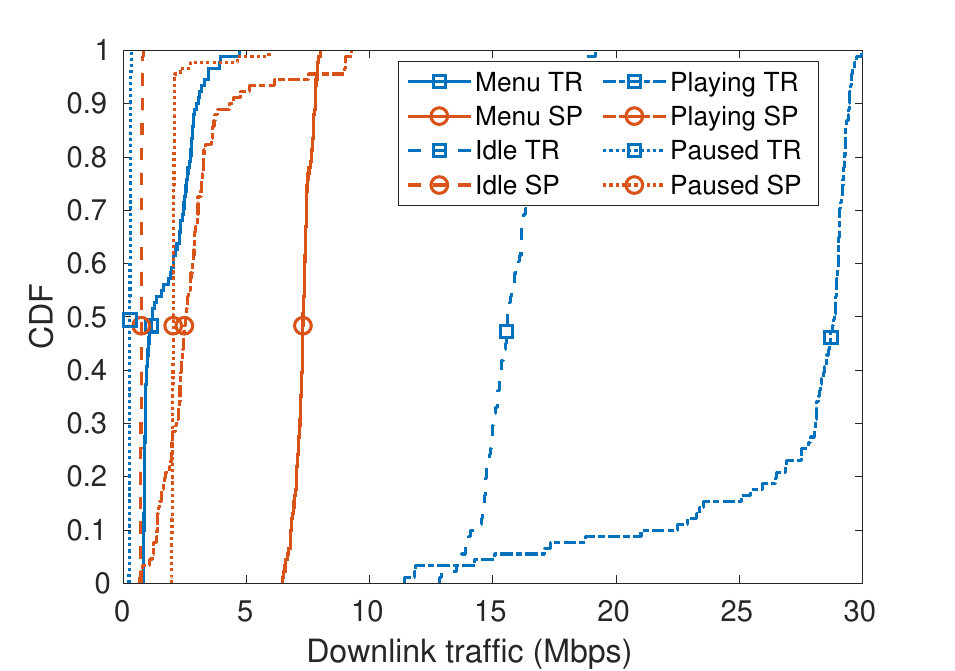}
    \caption{Traffic load}
    \label{traCDF}
\end{subfigure}
\begin{subfigure}[b]{0.329\textwidth}
\includegraphics[width=\textwidth]{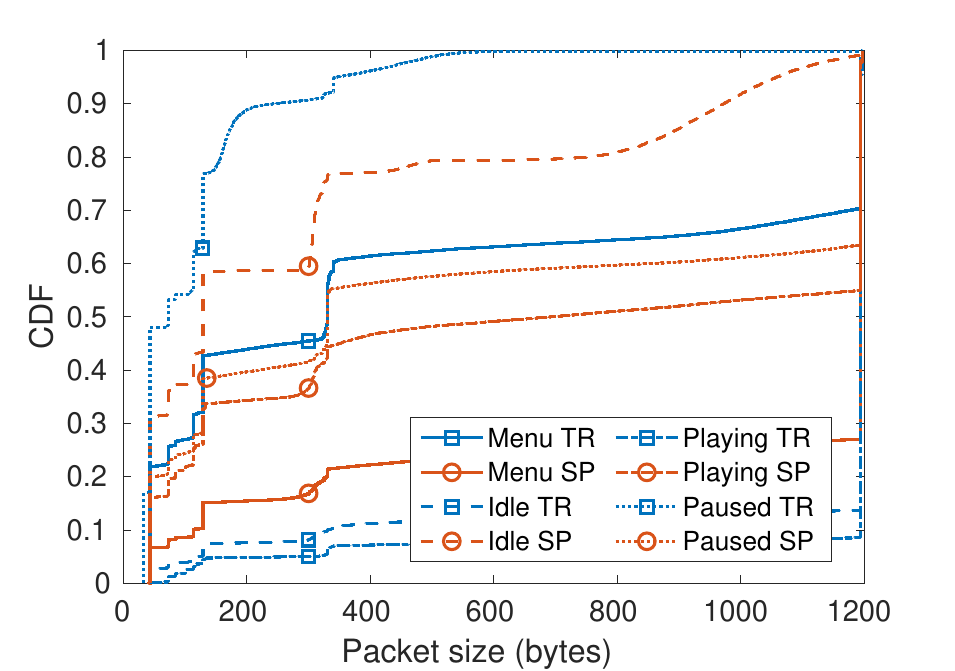}
    \caption{Packet size }
    \label{fsCDF}
\end{subfigure}
\begin{subfigure}[b]{0.329\textwidth}
\includegraphics[width=\textwidth]{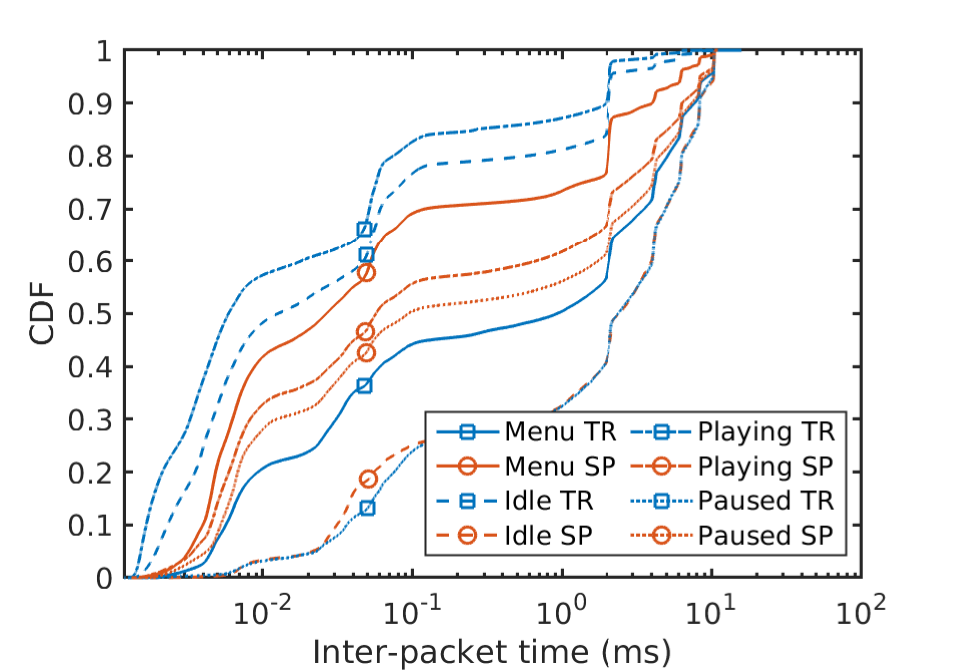}
\caption{Inter-packet time}
 \label{cdfs}
\end{subfigure}
\caption{CDFs for different games, and different game states inside a game.}
\label{two}
\end{figure*}

Figure~\ref{states} shows the traffic evolution for the different states of Tomb Raider and Spitlings, confirming that the traffic load depends on the current state. Note that the ``play" state is the only one where the player is pressing buttons frequently. If we compare the ``idle" and 'play' states, we can observe that there is a significant difference when the player interacts with the environment in terms of the traffic load. This is a result of the player's actions changing the information presented on screen, i.e., when ``idle" or in a 'menu', most of the elements on screen are static, and require no extra data. However, during playtime, new data needs to be sent to the user constantly. The lowest traffic load is found during the loading screen, as it is only a black background with a couple of lines of text. We can also observe that the traffic required per state is not similar across the games. Tomb Raider has a much higher throughput for the ``Idle" state than for the ``Menu", while the opposite is true for Spitlings. We also find that the variance in each state is not related to the average throughput. For example, in the 'main menu' of Tomb Raider we have an average traffic of 1.73 Mbps and a standard deviation of 0.96 Mbps, while in the ``idle" state, which has a  much higher average throughput of 15.79 Mbps, the standard deviation is only of 1.24 Mbps.

\textit{Finding}: Each game state has different traffic characteristics. Depending on the game, it could be the case that the ``play" state may not be the one with the highest traffic load, as it could be initially expected. While we expected some variance between states, some games can have more than 20 Mbps of difference between them, which could have quite an impact on a network's performance.


\subsection{Variability of the traffic load}

This section explores the traffic variability in each of the different game states. A higher variability should be expected in the ``play" state than in the other states due to the expected interaction with the user.

We use the same dataset as in the previous section (D3). For each game state of Tomb Raider and Spitlings we calculate the empirical cumulative density function using 90 seconds of data (we use the ``middle" section of each state, as to avoid interference from other states, such as the initial spikes in traffic when changing between states). We consider only the downlink RTP traffic, which includes the video and audio contents.

Figure~\ref{traCDF} shows the ecdf for each state, where we observe that the ``play" and ``idle" states of Tomb Raider result in the highest traffic load by far, reaching almost 30 Mbps and 20 Mbps, respectively. For Spitlings, as mentioned in previous section, the highest load appears in the main menu, which stays below 8 Mbps. All remaining states have relatively low loads, mostly staying below 5 Mbps.

Most states show little variability: the ``pause" state has a standard deviation of 0.02 Mbps for Tomb Raider and 0.49 Mbps for Spitlings; the main menu has 0.98 Mbps and 0.34 Mbps respectively. The ``idle" state has a deviation of 1.26 Mbps for Tomb Raider and 0.026 Mbps for Spitlings. The highest deviation appears in the ``play" state for both games, with 4.30 Mbps for Tomb Raider, and 1.72 Mbps for Spitlings, showing that the player actions have a clear impact on the generated traffic.

\textit{Finding}: The results confirm that states with high user interactions show the highest variance in their traffic, showing that the player actions have an impact on the generated traffic. 

\begin{figure}[ht]
    \centering
     \begin{subfigure}[b]{0.44\textwidth}
    \includegraphics[width = \textwidth]{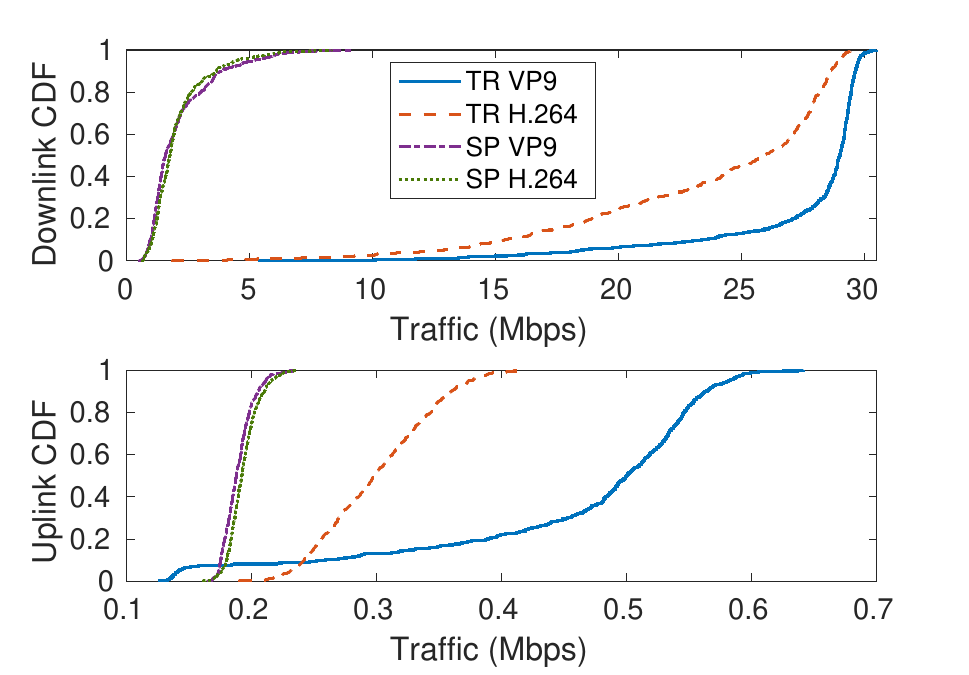}
    \caption{Traffic characteristics for H.264 and VP9 video codecs}
    \label{ThrCodecCDF}
    \end{subfigure}
    \begin{subfigure}[b]{0.44\textwidth}
    \includegraphics[width = \textwidth]{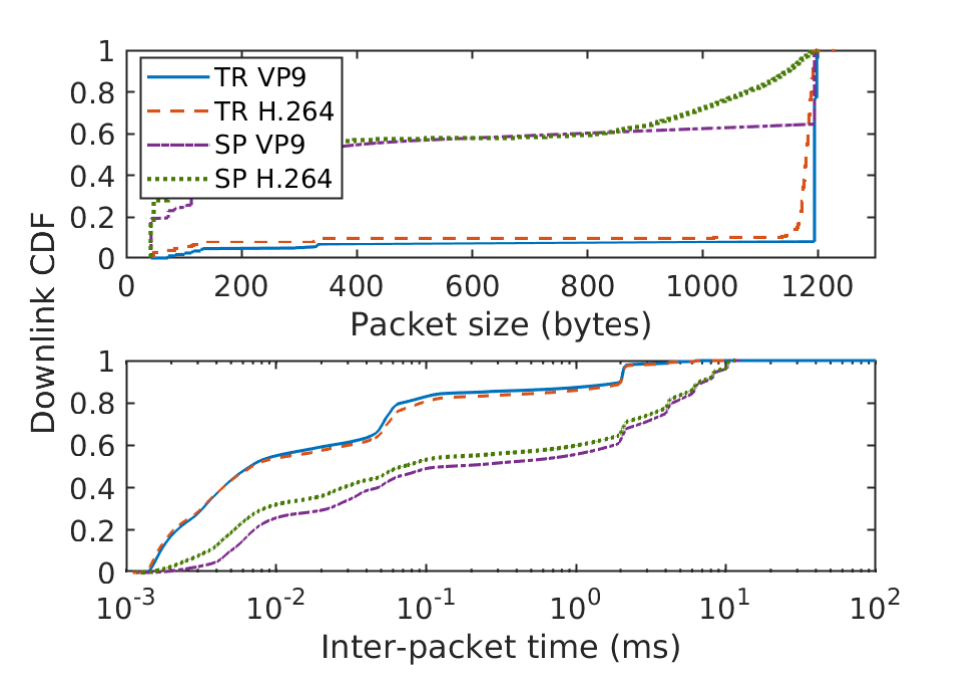}
    \caption{Packet size and inter-packet CDF for downlink traffic}
    \label{subplot}
    \end{subfigure}
    \caption{Video codec characteristics.}
\end{figure}


\section{Video Codecs and Resolution}\label{sec:video}

This section investigates the effect of the  video codec  and the video resolution on the traffic generated by Stadia.


\subsection{Codecs}

\begin{figure*}[t]
    \centering
     \begin{subfigure}[b]{0.329\textwidth}
    \includegraphics[width = \textwidth]{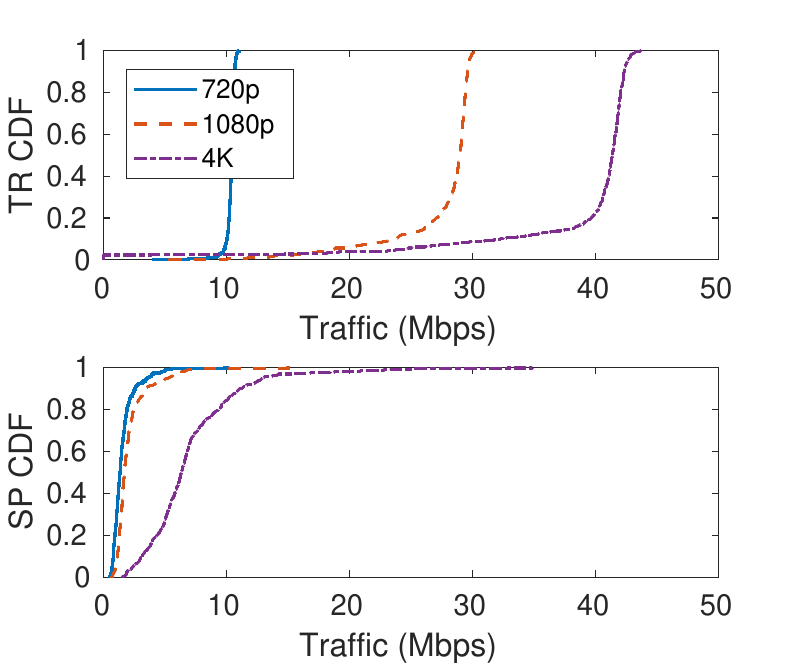}
    \caption{Traffic load}
\label{res1}
    \end{subfigure} 
    \begin{subfigure}[b]{0.329\textwidth}
    \includegraphics[width = \textwidth]{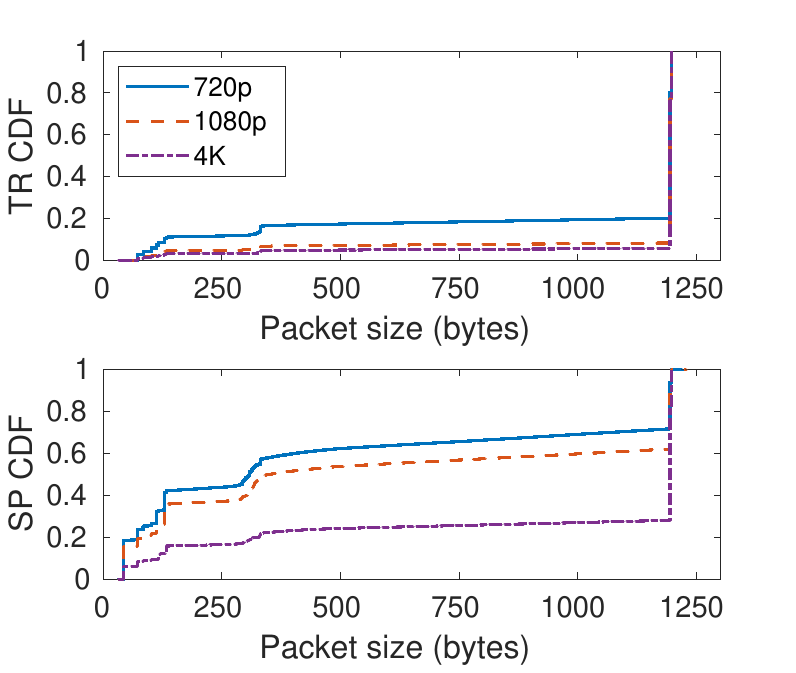}
    \caption{Packet size  }
    \label{res4}
    \end{subfigure}
    \begin{subfigure}[b]{0.329\textwidth}
    \includegraphics[width = \textwidth]{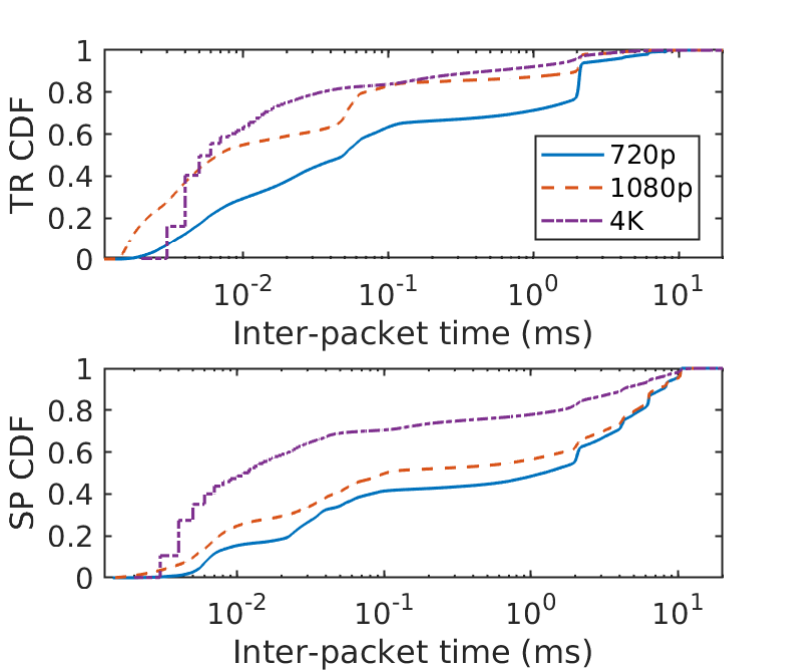}
    \caption{Inter-arrival time }
    \label{res6}
    \end{subfigure}
    \caption{Impact of the resolution on metrics.}
    \label{Fig:Figure9}
\end{figure*}

Google Stadia supports two different types of video encoding: VP9 and H.264. This section aims to identify if the use of different video codecs affects the generated Stadia traffic. The same parts of Tomb Raider and Spitlings  (``play" state) are played twice, one for each codec. Dataset D4 is used in this section.

Figure~\ref{ThrCodecCDF} shows the ecdf of the traffic load for both codecs. While Spitlings shows almost no differences between encodings, playing Tomb Raider using the H.264 codec represents less traffic. In average, the traffic load required for H.264 is 23.63 Mbps for Tomb Raider and 2.07 Mbps for Spitlings, and for VP9 it is 27.56 Mbps and 2.10 Mbps respectively. 

To dig a bit more on the differences between VP9 and H.264, we plot the downlink ecdf of the packet size and inter-packet time in Figure~\ref{subplot}. The average packet size using VP9 in the downlink is 1116.12 bytes and 530.04  bytes for Tomb Raider and Spitlings, respectively (in the uplink, the average is 75.71 bytes and 116.73 bytes). When the H.264 codec is used, the average packet size of downlink  packets is 1064.10  bytes and 480.34 bytes for Tomb Raider and Spitlings, with 123.83 and 118.24 bytes for the uplink. In general, VP9 attempts to use the same packet size whenever possible, which we can observe in the VP9 downlink of Tomb Raider, where 68.79\% of the packets are 1194 bytes long. Differently, for H.264, the most common packet size is 1183 bytes long, with 3.58\% of all packets being that size. Regarding inter-packet times, it can be observed that even if there are differences in the traffic load and packet size distributions, the two codecs have almost the same inter-packet time ecdf, which means that the traffic generation process is independent on the codec used.

\textit{Finding:} In our experiments, the use of H.264 codec has resulted in lower traffic loads than using VP9 for Tomb Raider. This is an unexpected result since VP9 is on paper a more advanced codec, and H.264 is supposed to be kept just for compatibility across all devices. Also, we can observe that the traffic generation process is exactly the same for both codecs, only slightly changing the packet size distribution, which in turn, results in different traffic loads. Although fully subjective, we did not experience any difference in terms of quality between the two video codecs.


\subsection{Resolution} \label{sec:resolution}

Stadia offers 3 different resolutions: 1280x720 (720p), 1920x1080 (1080p) and 3840x2160 (4K). Stadia recommends a minimum connection of 10 Mbps, 28 Mbps, and 35 Mbps to enjoy their service at 720p, 1080p, and 4K, respectively \cite{googleRec}. This section explores the relationship between the resolution and the traffic load. We also aim to validate if those capacity recommendations are accurate in practice. This section uses dataset D5, in which we play the same section of each game 3 times, one for each resolution. We use VP9 in all cases, and play the games for 600 seconds. 

Figure~\ref{res1} shows the ecdf of the Stadia traffic in the downlink for Tomb Raider and Spitlings. For Tomb Raider, we can observe that the ecdf for the 720p and 1080p resolutions closely follows the recommended link capacity. However, for the 4K resolution, the traffic load is higher than 35 Mbps for 88\% of the time, reaching up to 43.74 Mbps. For Spitlings, increasing the resolution from 720p to 1080p does not represent a large increase of the traffic generated, while for 4K, the increase is significant. For instance, considering the 95th percentile, the traffic load from 720p to 1080p increases in a 47.63\% (opposed to the 177.00\% for Tomb Raider), while from 1080p to 4K increases in a 144.15\% (only 43.09\% for Tomb Raider).

We have seen that using different resolutions results in a change on the traffic generated by Stadia. Different traffic loads are obtained by changing both the packet size distribution (Figure~\ref{res4}) and the inter-packet time distribution (Figure~\ref{res6}). First, regarding the packet size distribution, we can observe that reducing the resolution results in transmitting less packets over 1000 bytes. The number of packet groups inside each video frame period, as well as the number of packets in each group is also reduced. Second, as it could be expected, transmitting less packets affects also the inter-packet time distribution. However, we can observe that in spite of those differences, the shape of the ecdf is similar for all three resolutions, meaning that the traffic generation process follows the same general pattern regardless the resolution employed.

\begin{figure}[th!]
\centering
\includegraphics[width = 0.8\columnwidth]{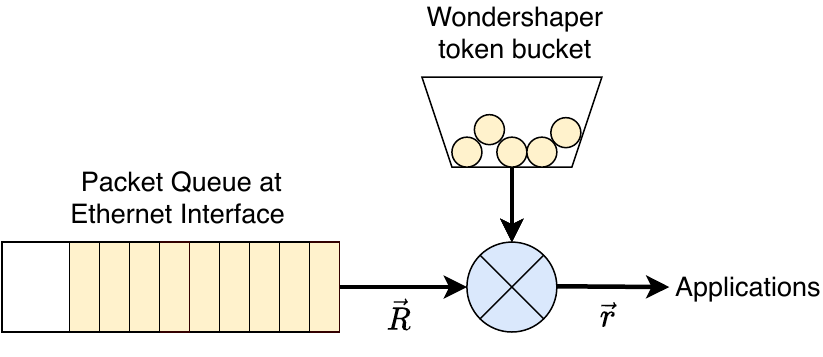}
\caption{Wondershaper operation: a token bucket allows new packets to get processed at a specified rate. }\label{wonders}
\end{figure}

\textit{Finding:} Supporting a resolution of 4K (i.e., up to 43 Mbps) requires a network capacity almost 4 times higher than for a resolution of 720p (which can go up to 11 Mbps). Moreover, the suggested link capacity values for each resolution are surpassed in all three cases. Changing the resolution affects the number of video packets generated, although the general packet generation process is unaffected. Subjectively, the use of higher resolutions is clearly observed in the quality of the image.


\section{How Stadia adapts to the available bandwidth}\label{sec:band}

This section studies the adaptability of Stadia to changing network conditions. We limit the available bandwidth at the client, and observe how performance is affected. We consider two cases: a) the game starts with the bandwidth limit already in place, and therefore, Stadia knows before starting the game the effective link capacity, and b) the available link capacity suddenly changes in the middle of the game, enabling us to observe how Stadia reacts.


\subsection{Different initial available bandwidths} \label{sec:initialbandlimits}

We start by applying a limit to the available bandwidth at the receiver before the game starts. Using different bandwidth limits will allow us to understand how Stadia deals with different link capacities. 

To limit the available bandwidth, we use Wondershaper\footnote{Wondershaper: \url{https://github.com/magnific0/wondershaper}}, a tool that uses Ubuntu's traffic control capabilities to limit incoming traffic. Wondershaper is installed on the receiving laptop, where it creates a virtual interface that receives incoming traffic and sends it to the physical interface following our specification. We use limits $\vec{r}$ of 5, 10, 15, 20, 30, 40 and 50 Mbps. For each limit, we play the same section of Tomb Raider for 60 seconds at 1080p with VP9. Figure~\ref{wonders} shows the operation of Wondershaper, and how it uses a token bucket to limit the network interface bandwidth $\vec{R}$ to the newly imposed $\vec{r}$. In this section we use dataset D6. 

\begin{figure*}[t]
    \centering
    \begin{subfigure}[b]{0.329\textwidth}
    \includegraphics[width = \textwidth]{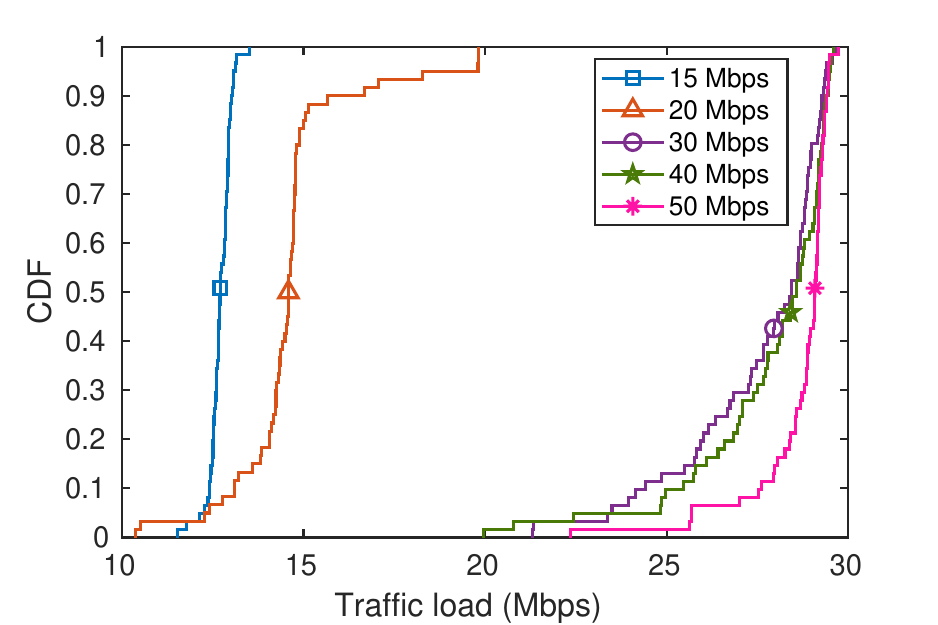}
    \caption{Downlink traffic CDF }
    \label{dt}
    \end{subfigure}
    \begin{subfigure}[b]{0.329\textwidth}
    \includegraphics[width = \textwidth]{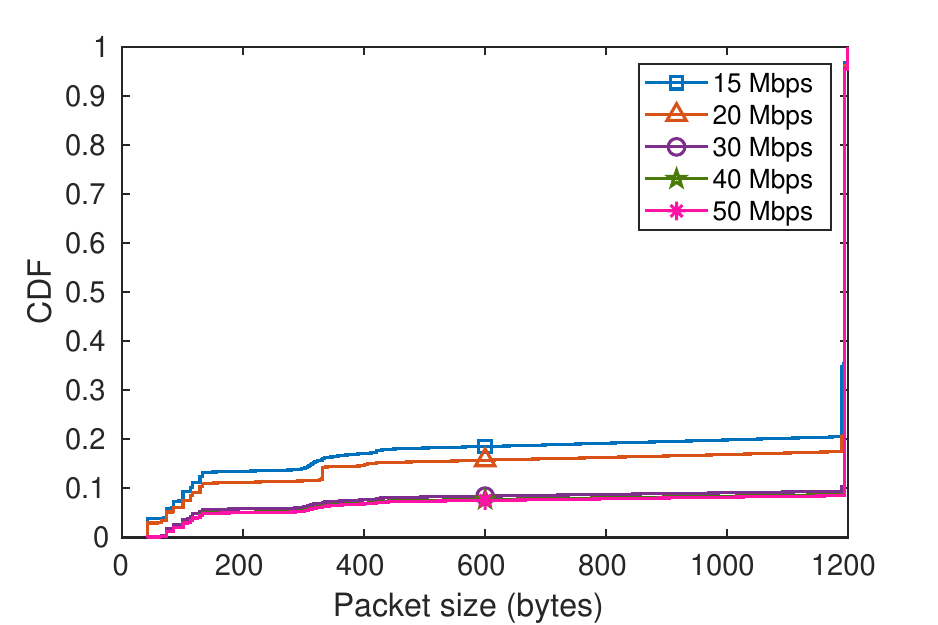}
    \caption{Packet size }
    \label{fs}
    \end{subfigure}
     \begin{subfigure}[b]{0.329\textwidth}
    \includegraphics[width = \textwidth]{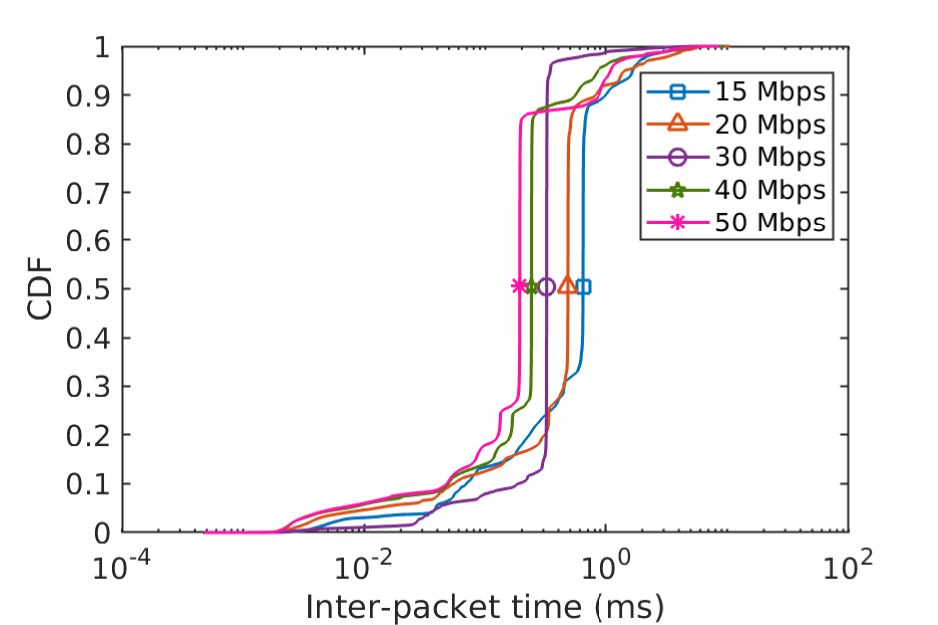}
    \caption{Inter-packet time }
    \label{if}
    \end{subfigure}
    \caption{Different link bandwidths.}
    \label{cdfscodecs}
\end{figure*}

\setlength\tabcolsep{10 pt}
\renewcommand{\arraystretch}{1.1}

	\begin{table*}[t]\centering
	\begin{small}
    \begin{tabular}{|c|c|c|c|c|c|c|}
  		\hline  
		\textbf{Limit}&  \textbf{Avg. packet size (bytes)} & \textbf{STDEV} &  \textbf{Min} & \textbf{Max} & \textbf{3 most common}\\ \hline
 
  15 Mbps & 995.61 & 405.42 & 43 & 1198 & 1194 (57.45\%), 1188 (13.38\%), 43 (3.66\%)\\\hline
  20 Mbps & 1026.61 & 379.14 &  43 & 1198  & 1194 (53.10\%), 1198 (2.75\%), 1188 (2.15\%) \\\hline
  30 Mbps & 1105.50 & 285.92 &  43 & 1198  & 1194 (78.78\%), 1198 (4.14\%), 73 (1.22\%) \\\hline
  40 Mbps & 1113.25 & 272.86 &  43 & 1198  & 1194 (80.34\%), 1198 (4.22\%), 73 (0.96\%) \\\hline
  50 Mbps & 1115.21 & 269.84 &  43 & 1198  & 1194 (83.78\%), 1198 (4.40\%), 73 (1.01\%) \\\hline
  	
	\end{tabular}	\caption{Packet characteristics. }
	\label{downloads0}
	\end{small}
	\end{table*}
Figure~\ref{dt} shows the ecdf of the traffic load for each bandwidth limit. We can observe that Wondershaper guarantees that the traffic streams will use a bandwidth lower than the limit in average. For example, for bandwidths limits of 15 Mbps and 20 Mbps, Stadia generates a mean traffic load of 12.71 Mbps and 14.64 Mbps, respectively. For 30, 40, and 50 Mbps, the traffic generated by Stadia is almost the same in all the three cases. The mean traffic load values are 27.54 Mbps, 27.80 Mbps and 28.66 Mbps for each bandwidth limit, respectively. In addition, we can also observe that the highest traffic peak is never higher than 30 Mbps, even when bandwidth limits of 40 and 50 Mbps are in use.

Figure~\ref{fs} shows that the packet size distribution is almost identical in all cases regardless the imposed bandwidth limit. For 15 and 20 Mbps, there is a higher amount of small packets, but overall, the results are very similar to the cases of 30, 40 and 50 Mbps. Table \ref{downloads0} shows the average packet size and standard deviation for each case. We can observe how the average packet size increases along with the bandwidth limit, but the standard deviation decreases, as most of the packets are of the same size. We also show the most common packet sizes, where we can see the clear preference for packets of 1194 bytes no matter the bandwidth limit used.

Similarly to the packet size, the distribution of inter-packet times (Figure~\ref{if}) shows an identical tendency in all cases, with the highest bandwidth limits leading to slightly lower inter-packet times. 
Stadia already sustained 1080p with the 20 Mbps limit, only switching to 720p for a link capacity of 15 Mbps. This leads us to conjecture that Stadia can further stress the video encoding to reduce the traffic load, changing the resolution only as a last resort. We also used limits of 5 Mbps and 10 Mbps, but these low bandwidth values lead Stadia to stop the game shortly before starting, showing a message informing that the network was unsuited for the service.

\textit{Finding:} Stadia adapts to the available bandwidth seamlessly when the limitation is set before the game starts. We have found that Stadia strives to keep the 1080p resolution at 60 fps even if the available bandwidth is far below its own pre-defined requirements, and only switches to a lower resolution of 720p as the last resort. In this regard, subjectively speaking, the more compressed 1080p streams were still preferable than the 720p streams, justifying Stadia's behavior.


\subsection{Sudden changes on the available bandwidth} \label{sec:suddenchanges}

Here, we test the ability of Stadia to adapt to a sudden change in the available bandwidth. Limiting the bandwidth during the gameplay, we aim to investigate how Stadia responds to congestion in real time.

This section uses dataset D7. We start Wondershaper in the background after 120 seconds of playing without any bandwidth limit (i.e., using the default link capacity of 100 Mbps). We first consider bandwidth drops to 10 Mbps, 15 Mbps, 20 Mbps and 30 Mbps. After that, we do the opposite, we start Stadia under a bandwidth limit (as in the previous section), to remove it after 120 seconds. To showcase the impact of these changes, we use the metrics obtained from \url{chrome://webrtc-internals/}, such as the RTT (calculated as the time between the transmission of STUN packets and the arrival of the response), video packet losses and the rate of successful delivered video frames to the application. The use of Chrome's statistics to quantify QoS was studied in \cite{ammar2016video}.

\begin{figure*}
\centering
\begin{subfigure}[b]{0.28\textwidth}
\includegraphics[width = \textwidth]{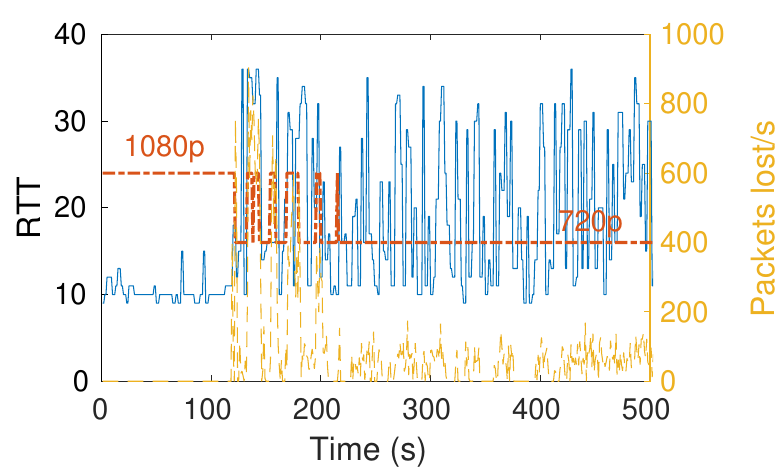}
\caption{Drop to 10 Mbps}
\label{rtt1}
\end{subfigure}
\begin{subfigure}[b]{0.28\textwidth}
\includegraphics[width = \textwidth]{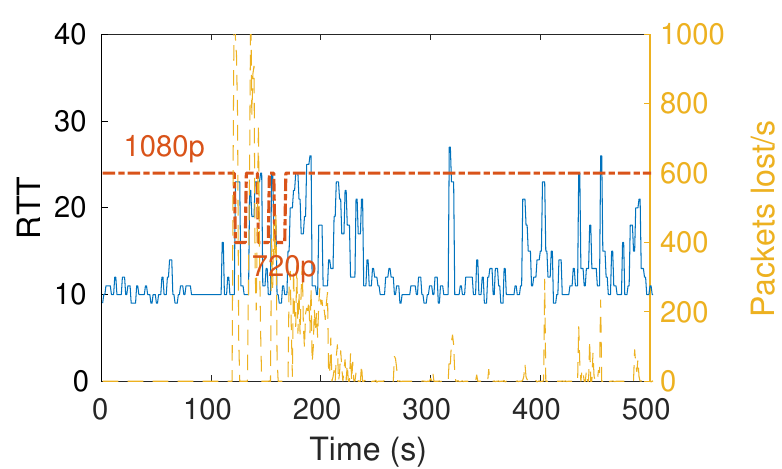}
\caption{Drop to 20 Mbps}\label{rtt2}
\end{subfigure}
\begin{subfigure}[b]{0.28\textwidth}
\includegraphics[width = \textwidth]{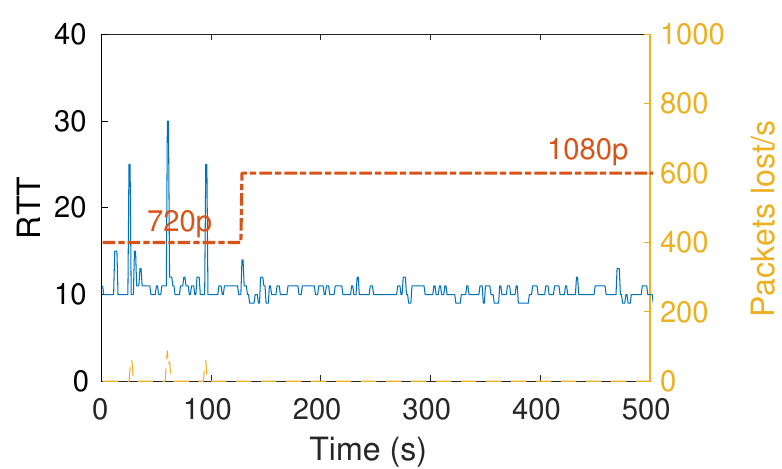}
\caption{15 Mbps to 100 Mbps }\label{jit2}
\end{subfigure}
\begin{subfigure}[b]{0.28\textwidth}
\includegraphics[width = \textwidth]{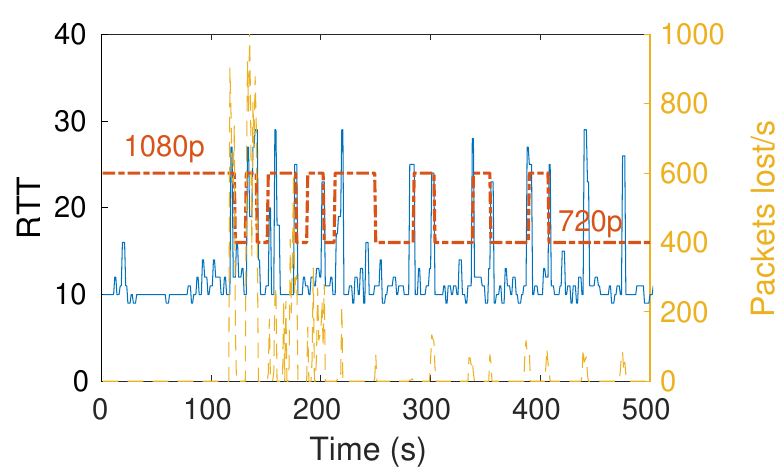}
\caption{Drop to 15 Mbps }\label{jit1}
\end{subfigure}
\begin{subfigure}[b]{0.28\textwidth}
\includegraphics[width = \textwidth]{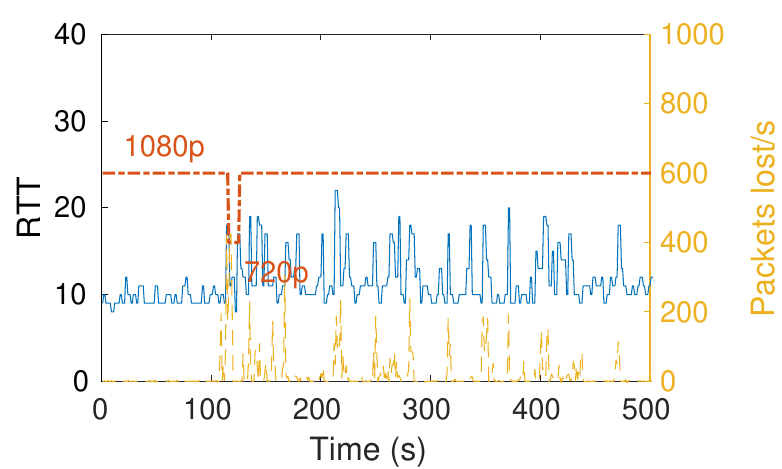}
\caption{Drop to 30 Mbps}\label{jit3}
\end{subfigure}
\begin{subfigure}[b]{0.28\textwidth}
\includegraphics[width = \textwidth]{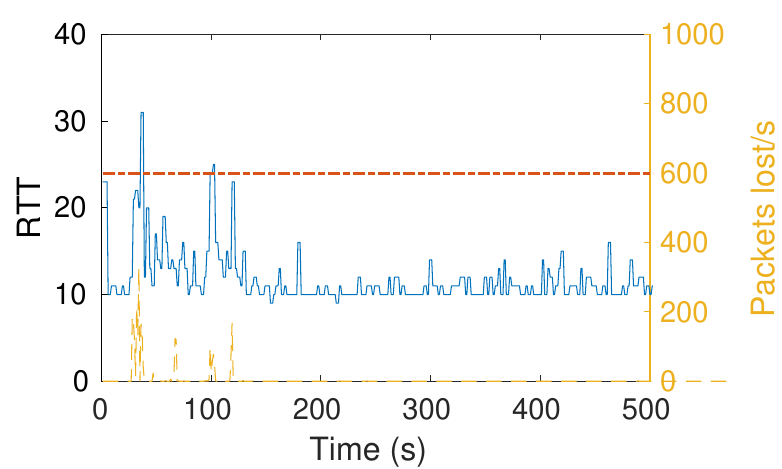}
\caption{20 to 100 Mbps}\label{jit4}
\end{subfigure}
\caption{Round Trip Time (continuous line), video packets lost (dashed line) and resolution (dash-dotted line). }
\label{ada}
\end{figure*}

Figure~\ref{ada} shows the RTT, video packets lost and video resolution of the stream over time for each of the bandwidth limits. We can observe that once the bandwidth decreases, Stadia  changes resolution repeatedly for a while. Once the client reports that packets are being lost, Stadia drops to the lowest resolution. Then, after a short time, Stadia attempts to use a new encoding configuration with a higher resolution. If the new configuration still results in dropping packets, Stadia repeats the process again until a viable configuration is found. 

A change in the available bandwidth leads to transitory periods of variable duration during which Stadia attempts to recover by finding a viable configuration. The bandwidth drop to 10 Mbps results in Stadia changing resolutions 12 times during 95 seconds before remaining at 720p until the end. It is worth to mention here that this game session was completed successfully, while in the previous section, starting with the limit of 10 Mbps already in place resulted in Stadia deciding to close the game. The bandwidth drop to 15 Mbps in Figure~\ref{jit1} leads to the longest transitory period, spanning 287 seconds before Stadia stops changing resolutions and remains at 720p. Dropping the available bandwidth to 20 Mbps results in a comparatively quick reconfiguration of 47 seconds, and dropping it to 30 Mbps results in a single switch to 720p for 11 seconds before returning to 1080p. 

When we start Stadia with a bandwidth limit and then remove it, the time to find a new viable configuration is much faster, as it could be expected. For the case where the initial limit was 15 Mbps, Stadia just jumps to 1080p. In the case that the initial bandwidth limit was 20 Mbps, Stadia never changed the resolution, as we started at 1080p already, but the generated traffic load does increase, going from 17.75 Mbps before removing the limit, to 24.71 Mbps afterwards, further confirming the existence of different coding settings for the same resolution. 

The RTT increases rapidly when the bandwidth limits are enforced, going from an average of 10 ms for most captures to 19.52 ms for the bandwidth drop to 10 Mbps. The variance in the RTT is higher for the lower bandwidth limits, as we can observe peaks of more than 35 ms in Figure~\ref{rtt1}, while we only reach 22 ms for 30 Mbps in Figure~\ref{jit3}. We can also observe an increase in the packets lost when the bandwidth limits are enforced, reaching as far as 1017 packets lost in a single second in Figure~\ref{rtt2}. We can observe that the packets lost decrease during the transitory period in Figures \ref{rtt1} and  \ref{jit1}, stabilizing afterwards. For the two cases in which the bandwidth limit is removed, we observe that the RTT stabilizes soon at 10 ms, and Stadia stops dropping packets, showing the opposite tendency to the previous cases.

\begin{figure}
    \centering
    \includegraphics[width=0.40\textwidth]{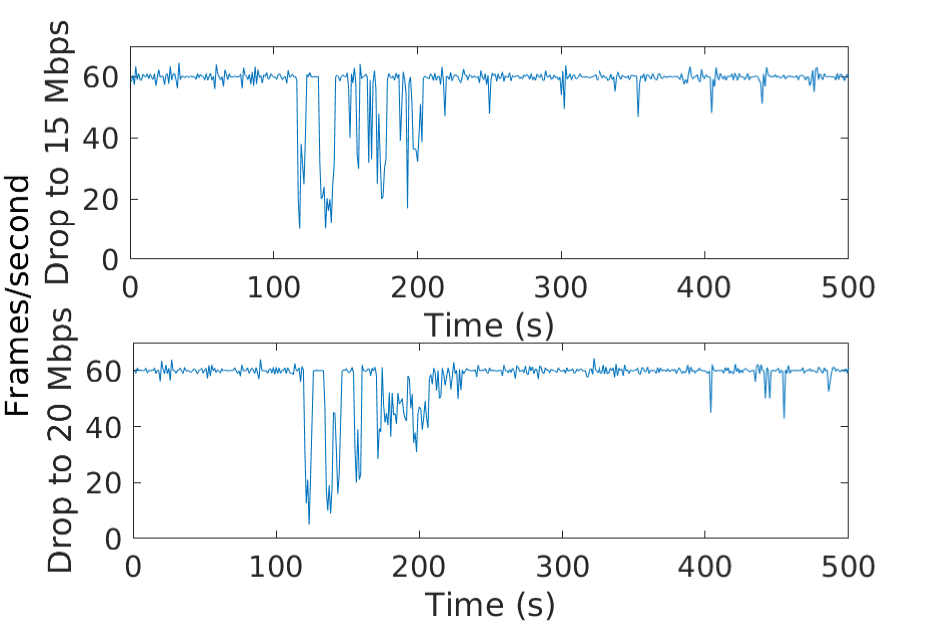}
    \caption{Framerate over time.}
    \label{fpsovertime}
\end{figure}

\begin{figure*}
\centering
\begin{subfigure}[b]{0.329\textwidth}
\includegraphics[width = \textwidth]{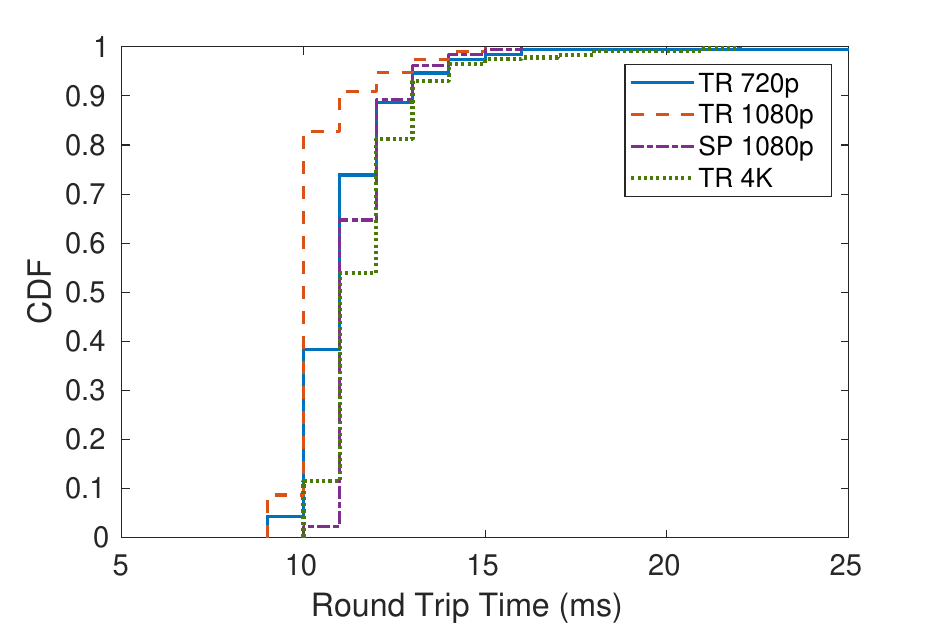}
\caption{RTT for each game}
\label{rtt10}
\end{subfigure}
\begin{subfigure}[b]{0.329\textwidth}
\includegraphics[width = \textwidth]{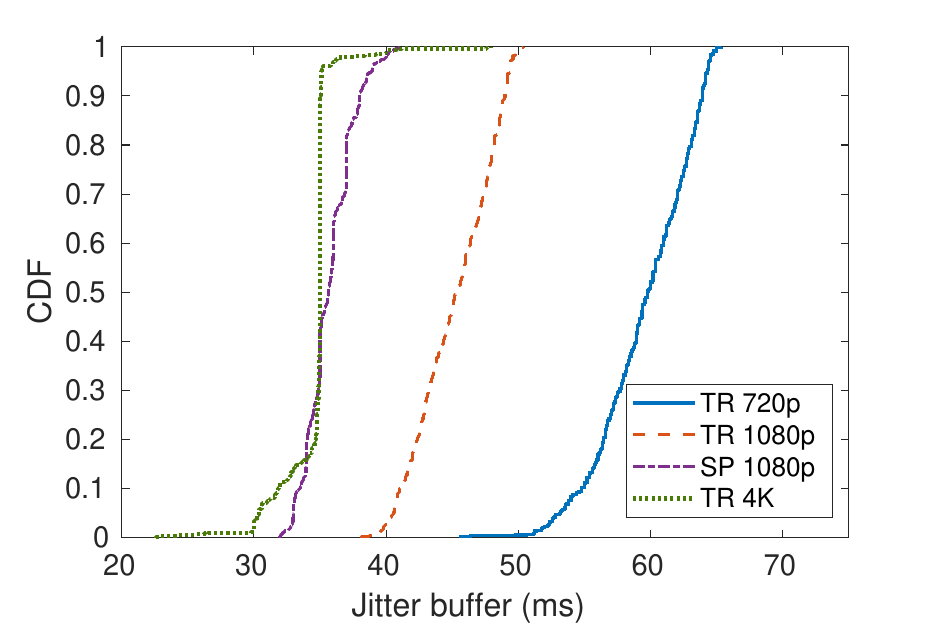}
\caption{Jitter buffer delay}\label{jit10}
\end{subfigure}
\begin{subfigure}[b]{0.329\textwidth}
\includegraphics[width = \textwidth]{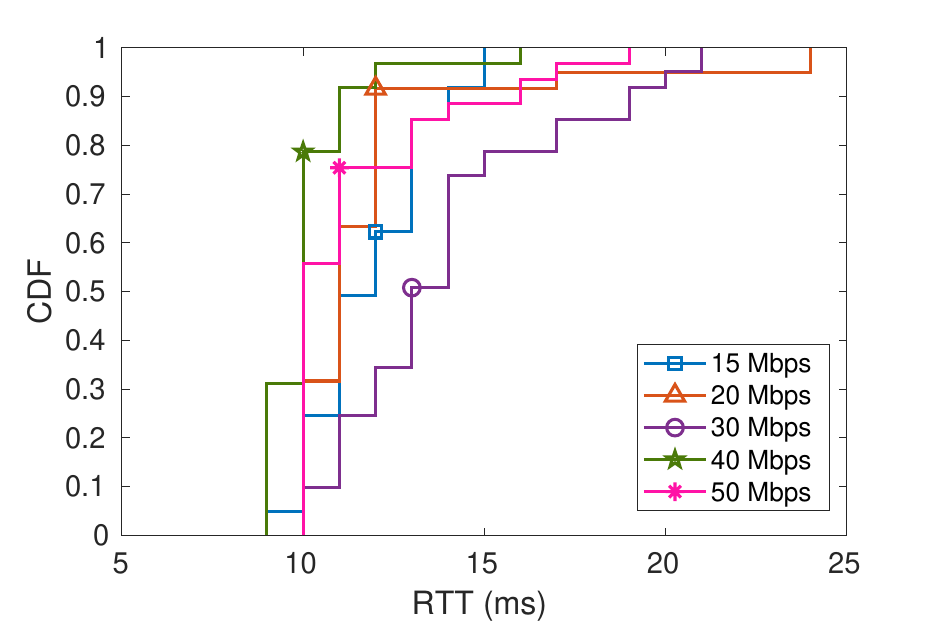}
\caption{RTT when bandwidth is limited}\label{rttbw}
\end{subfigure}
\caption{Round Trip Time and jitter buffer delay for different configurations.}
\label{ada2}
\end{figure*}

Figure~\ref{fpsovertime} shows the video frames per second decoded by the browser over time for the bandwidth drops to 15 and 20 Mbps. Correct playback should lead to a stable 60 fps. However, we can observe that during the transitory period the framerate varies strongly, dropping as low as 10 fps  for 15 Mbps, and 5 fps for 20 Mbps. Combined with the changes in the resolution (see Figure~\ref{res1} for the impact of resolution on traffic), the reduction on the framerate results in a noticeable loss of quality for the user. While the resolution has an effect on the quality of the image, the impact of the framerate is on the smoothness of the video reproduction. Frame drops such as these found in Figure~\ref{fpsovertime} result in severe delays between the time the player presses a button and the time the corresponding action appears in the screen, as well as in parts of the video being skipped, resulting in characters  ``teleporting" to another place due to the missing frames. Finally, let us mention that in the previous case, when Stadia started with a bandwidth limit already in place, the framerate remains stable for the entire capture, so the user notices the low bandwidth availability only in the image resolution. 

\textit{Finding:} Stadia attempts to recover from a drop in the available link capacity almost immediately, entering a transient phase in which Stadia aims to find a new configuration to compensate the lack of network resources. This transient phase however, can last over 200 seconds. During this time, although the user is able to continue playing, the quality of experience is heavily affected, with constant resolution changes, inconsistent framerate and even audio stuttering. In the case that the available bandwidth increases, Stadia also reacts immediately, easily finding a new viable configuration.

\section{Latency related to bandwidth changes}\label{sec:latency}

In most of previous experiments we have intentionally avoided any reference to the end-to-end latency, so we can focus on it in this and the following sections. In this one, we analyse the latency of our previous experiments, i.e., latency under normal conditions and then with sudden drops in link capacity. We use the RTT and jitter buffer delay metrics from WebRTC internals (\url{chrome://webrtc-internals}). The RTT is computed as the total time elapsed between the most recent STUN request and its response, and it is reported every second. It  shows how long it takes for an action to obtain a response. The jitter buffer delay represents the amount of time RTP packets are further buffered at the client side to guarantee a smooth data delivery to the user (i.e., fixing packet order, since UDP does not offer such control by itself). Note that the jitter buffer delay is adaptive and so it may change with time.

High latency can have a negative impact on player enjoyment. After pressing a button, players expect that the effects of the corresponding action will appear instantaneously on the screen. If it takes too long, it can make the game unenjoyable, and in some cases, fully disrupt the gameplay. Here, we compare the RTT and jitter buffer delay metrics for different games and resolutions.

Dataset D8 presents the WebRTC internals that correspond to the same traffic captures as in Sections \ref{sec:resolution} and \ref{sec:initialbandlimits}.

Figure~\ref{rtt10} shows the ecdf of the RTT for both Tomb Raider and Spitlings during the ``play" state at 1080p, as well as the RTT for Tomb Raider at 720p and 4K. All of them show similar RTT values, averaging between 10.28 ms and 12.30~ms. In all cases, the 95 \% percentile of the RTT values is lower than the duration of a single video frame (16.67 ms), meaning that in these tests Stadia had the opportunity to interact with the player's actions without any perceptible delay. As expected, since the available link capacity was large enough, the game and resolution have no impact on the perceived latency. Only when the traffic load is close to the network capacity, such as in the previous section, the RTT is affected.

Figure~\ref{jit10} shows the empirical ecdf of the jitter buffer delay for Tomb Raider and Spitlings. For a resolution of 1080p, Spitlings shows a lower jitter buffer than Tomb Raider,  with an average of 35.76 ms and 45.35 ms, respectively, which corresponds to 2-3 video frames at a framerate of 60 fps. Considering Tomb Raider and different resolutions, we observe that the jitter buffer decreases as the resolution increases. For Tomb Raider the jitter buffer averages 58.42 ms, 45.34 ms and 35.35 ms for 720p, 1080p and 4K, respectively. This is an interesting result that shows the buffer jitter delay is directly related to the inter-packet arrival time (i.e., lower inter-packet arrival times also result in lower jiter buffer values, and the opposite is also true).

Figure~\ref{rttbw} shows the RTT for the different initial bandwidth limits considered in Section \ref{sec:initialbandlimits}. We can observe that the RTT is quite similar to the ones on Figure \ref{rtt10}, showing that our bandwidth limitations do not have an impact on the RTT, and only affect the throughput of the network.

\textit{Finding:} In all the considered cases for different games, resolutions, and available bandwidth limits, RTT values have consistently been below 25 ms, with average values between 10 and 15 ms. This is especially relevant since all the experiments and measurements have been done in a temporal span of several months from March to July 2020, hence, meaning that in absence of network congestion, RTTs are extremely stable and clearly below 60 ms (i.e., the value at which users start noticing the latency issues  \cite{outatime}). In terms of the jitter buffer, we have found that it is higher for lower resolutions, as it directly depends on the amount of traffic received at the client.


\begin{figure*}[ht!]
\centering 
\begin{subfigure}[b]{0.32\textwidth}
\includegraphics[width=\textwidth]{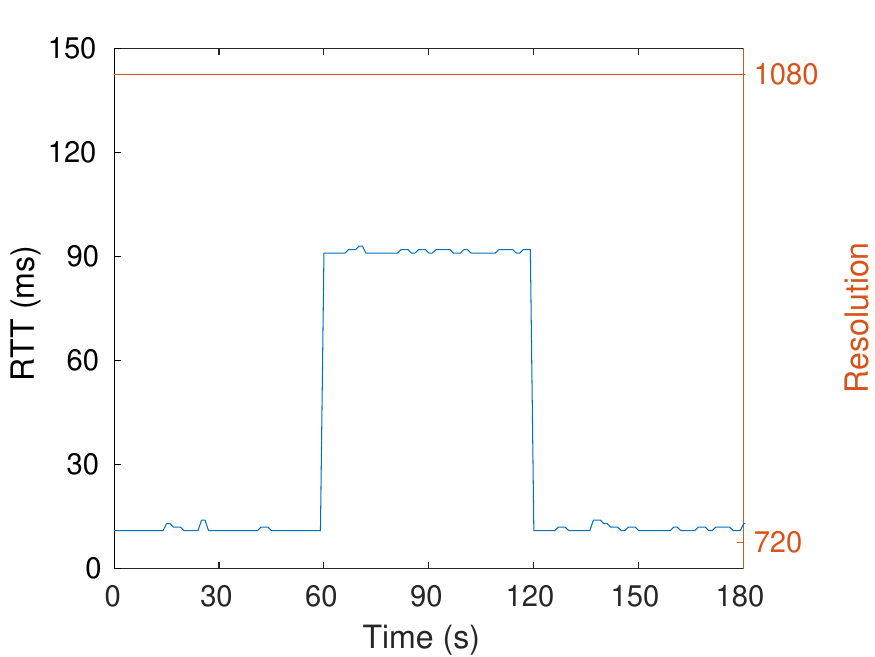}
\caption{ RTT increase of 80 ms}
 \label{ft}
\end{subfigure}
\begin{subfigure}[b]{0.32\textwidth}
\includegraphics[width=\textwidth]{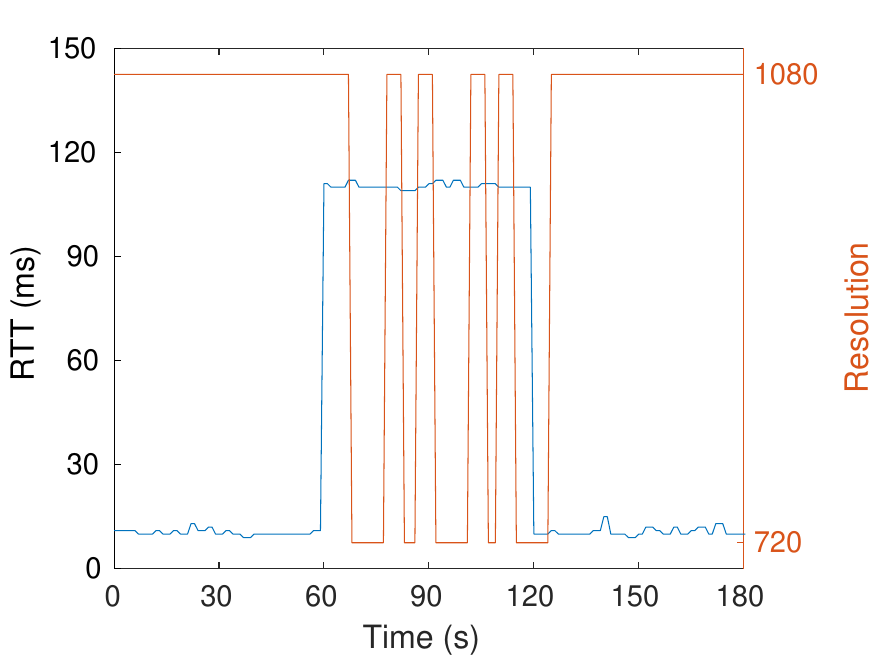}
    \caption{RTT increase of 100 ms}
    \label{ff}
\end{subfigure}
\begin{subfigure}[b]{0.32\textwidth}
\includegraphics[width=\textwidth]{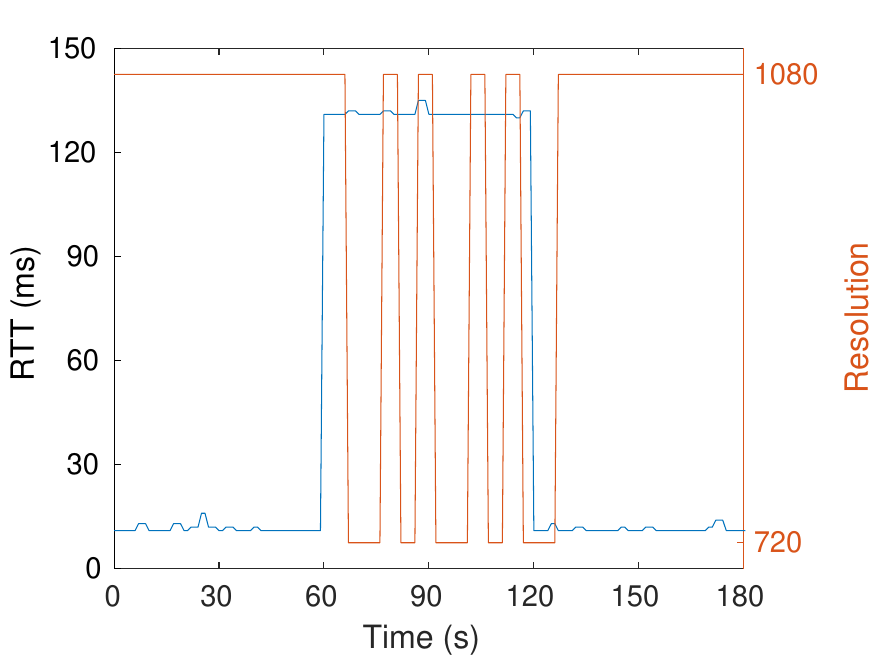}
    \caption{RTT increase of 120 ms}
    \label{st}
\end{subfigure}
\caption{Round Trip Time and video resolution over time.}
\label{tt}
\end{figure*}



\section{How Stadia reacts to latency changes}\label{sec:latencychanges}

In this section we directly modify the latency in our network to investigate how Stadia traffic adapts to changes in the delay between packets. We use the Ubuntu Traffic Control (tc) package to change the latency of both incoming and outgoing packets in our PC, and much like before, we use the WebRTC internals metrics from Chrome to assess how latency changes impact the stream performance.

We start playing the game without any added latency. Then, after $60$ seconds, we add the same extra latency value to both incoming and outgoing packets from the client. We add $40$ ms, $50$ ms and $60$ ms to each link (i.e., a perceived addition of $80$ ms, $100$ ms and $120$ ms to the Round Trip Time). Then, after 60 seconds of high latency, it is removed for another 60 seconds.

Figure \ref{tt} shows the RTT of Stadia traffic over time for Tomb Raider, as well as the video resolution. It can be observed that for an extra RTT latency of $80$ ms the video resolution does not change. However, when we add $100$ and $120$ ms, the video resolution starts to go up and down, showing that Stadia reacts to the perceived congestion by trying different video configurations. 

Google Congestion Control works based on the bitrate, packet loss and delay of transmissions as explained in Section \ref{gccex}. Increasing the latency does not cause packet losses, and so changes in the configuration come purely from the delay-based controller. Once the latency goes back down, after 120 s, the system returns to the initial configuration quickly. In all cases, Stadia shows a message warning the user that their network is unstable after around 40 seconds have passed since the increase of latency. When the latency added is higher than 80 ms, if it is kept for more than $60$ seconds, the system would sometimes shut down on its own, giving the user another message explaining the situation. Gameplay continued uninterrupted for the duration of the experiment, with additional lag being noticeable, but not unmanageable. Framerate was consistent around 60 fps through the entire time, with few video frames being dropped and only occasional stuttering. As mentioned before, Chrome reported no packets lost either. In terms of player experience, the changes in bandwidth were a much bigger issue than the latency ones, as long as Stadia does not terminate the session by itself. 

\begin{figure}
    \centering
    \includegraphics[width=0.45\textwidth]{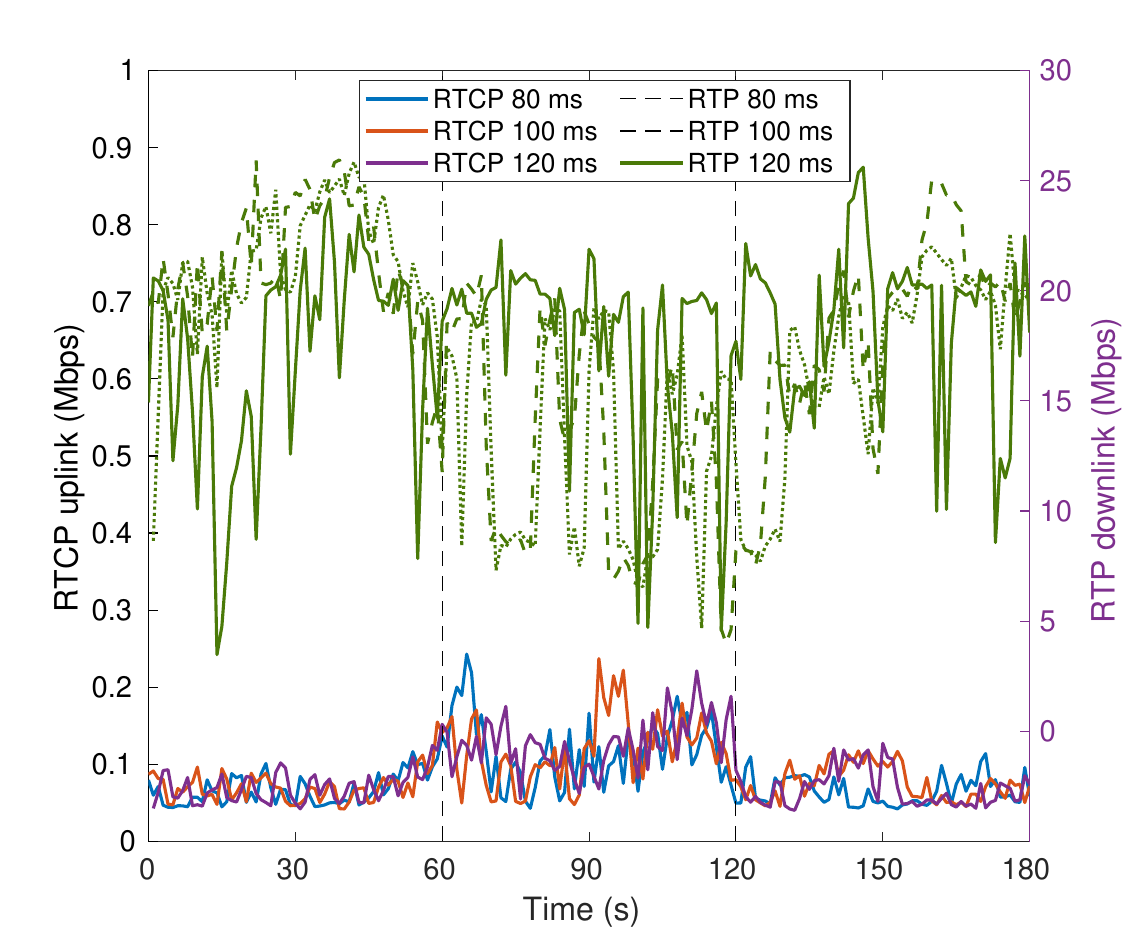}
    \caption{Impact of latency changes on the throughput.}
    \label{lat}
\end{figure}

Figure \ref{lat} shows the impact of the latency on the data exchanged with the Stadia server. An increase in uplink data can be observed during the period going from 60 seconds to 120 seconds (i.e., when the latency is increased), which then is reduced when latency returns to normal. We can also observe a decrease in downlink RTP video for all cases on the high latency period, including the case of 80 ms in which resolution did not change.  In the 80 ms case, it seems that the changes to the configuration are less aggressive, confirming that much like with bandwidth, changes in Stadia configuration are not just based on a binary overloaded/underloaded approach and also continuous adjustments to the video codec are applied based on the gathered feedback.

\textit{Finding:} Stadia reacts to latency changes quickly, increasing the client-side reports and adapting the video configuration (including video resolution) to reduce downlink traffic. Warnings are given to the client if the RTT is considered too high. After a few minutes of high latency (over 70 ms RTT) the connection is considered unsustainable, and the game is stopped.

\begin{figure*}[ht!!!]
    \centering
    \includegraphics[width = \textwidth]{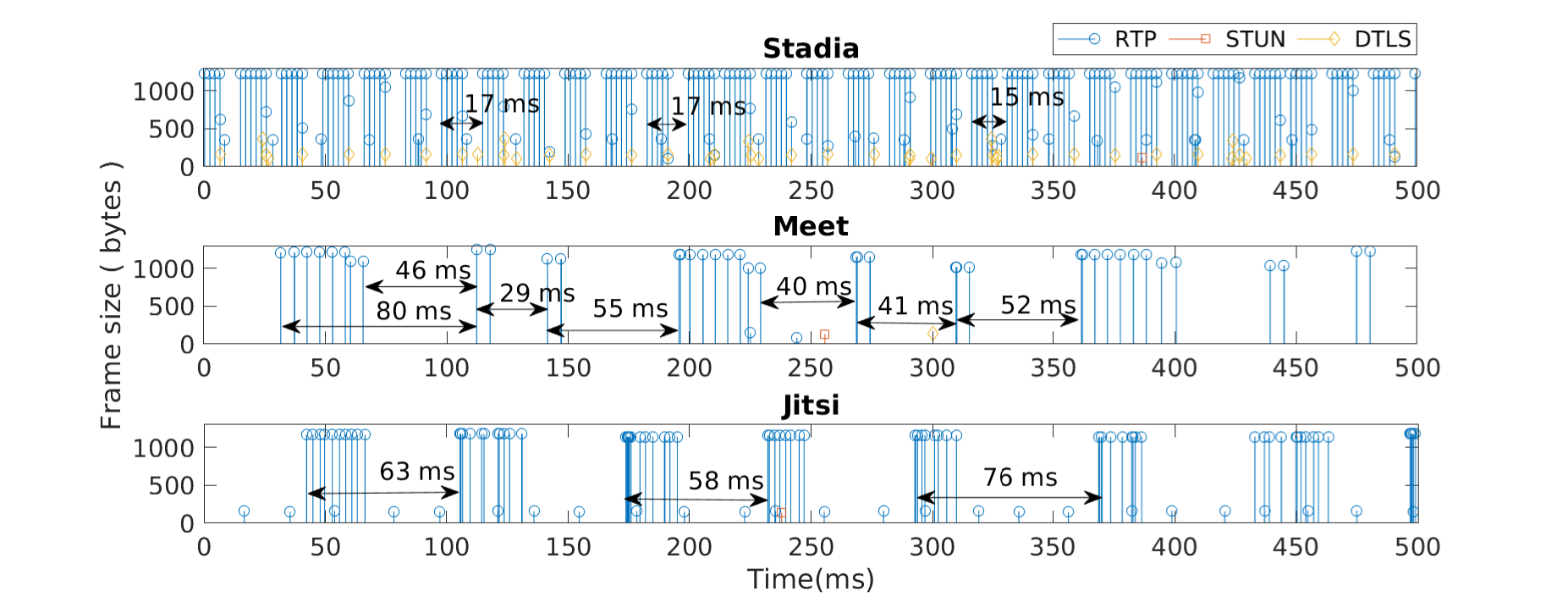}
    \caption{Temporal patterns of downlink traffic for WebRTC applications. }
    \label{groupingsWebRTC}
\end{figure*}

\section{Is Stadia traffic similar to other WebRTC applications?}\label{sec:otherWebRTC}

Stadia traffic is comprised mostly of video, and so we want to study which parts of the traffic are unique to Stadia and which are just part of a standard video application. As Stadia uses WebRTC, we will also compare its traffic to some video conferencing applications that use it, such as Google meet.

For this section we perform captures with other video services. We use the same setup we used for Stadia, but with Google meet\footnote{\url{https://meet.google.com/}} and Jitsi\footnote{\url{https://meet.jit.si/}} and a remote caller. Both of these applications are based on WebRTC, so we expect their traffic to have some similarities to that of Stadia. For both applications, both users use the highest video settings available, which means that the streams are 720p. The video codec used by Google Meet is VP9, just like Stadia. Jitsi however used VP8 in our captures. 
\begin{figure*}[ht]
    \centering
    \includegraphics[width = \textwidth]{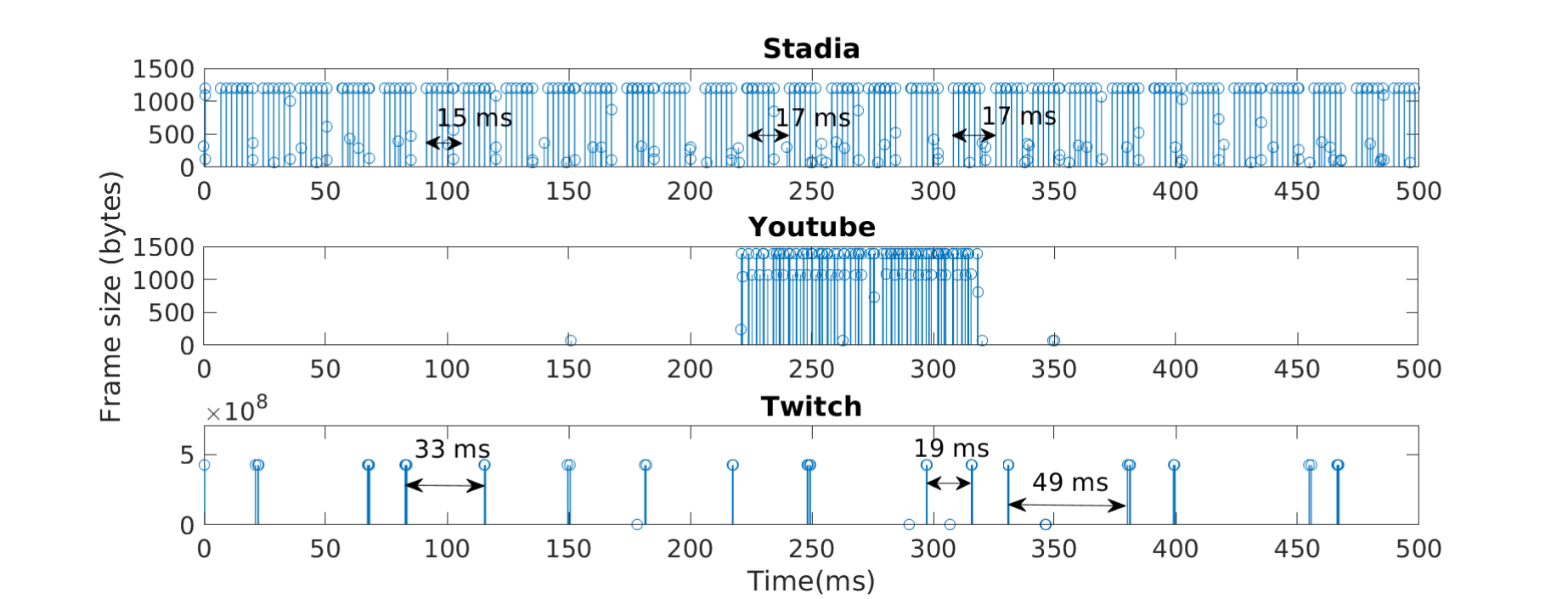}
    \caption{Temporal patterns of downlink traffic for high definition live video applications.}
    \label{groupingsvideo}
\end{figure*}
 To compare Stadia with the conferencing apps, we use a 720p capture of Tomb Raider so that all captures have the same resolution. In terms of framerate, Stadia is the only one that uses 60 fps, while Google meet had an average of 24 fps, and Jitsi used 15 fps \footnote{ Jitsi can go up to 30 fps depending on settings, but in our HD captures it settled around 15-20 fps.}. 
Figure \ref{groupingsWebRTC} shows the downlink traffic patterns of the three WebRTC applications we use over a period of 500 ms. For Stadia we can find the same patterns we described in Section \ref{sec:und_Stadia_traffic}, mainly that RTP packets arrive in batches, separated by around 16.67 ms, which is the time between two video frames at 60 fps. This is the main part of the traffic, with STUN and DTLS packets happening less frequently, and with small sizes.

For Jitsi we find that the patterns are very similar, with RTP packets arriving in batches, but with much larger intervals. A framerate of 15 fps would result in video frames lasting an average of 66.7 ms, which is consistent with the timing we observe. For Google meet however the RTP patterns are a bit different, where we can observe a large batch of 6 groups of packets, with two extra packets at the end that are smaller in size, followed by two small batches of 2 groups of packets. The timing for an average of 24 fps would be 41.7 ms, and we cannot identify when video frames start as easily as with the other two applications. The time between the bigger batches and the smaller ones is quite large at 80 ms, but the time between the last packet of the batch and the first of the next one is 46 ms, which fits much better with the framerate. The timing between the smaller batches also seems to fit closer to the average of 41.7 ms. Overall, while there are some differences between the traffic shape of meet and the other apps, the underlying patterns are quite similar, with large RTP packets sent in batches with intervals related to the video framerate.

Table \ref{supertaula2} summarizes other traffic parameters such as average packet size, inter packet time and traffic load. Average packet size is consistent across the three applications, as RTP traffic has the largest packets, while STUN and DTLS have much smaller ones. As a consequence of the lower framerates, we can observe that the average inter packet time is much larger for Meet and Jitsi, both 5 times higher than that of Stadia. In terms of RTP load we can clearly observe that Meet and Jitsi are incredibly close, both at 1.6 Mbps, while Stadia requires a much larger load of 10.54 Mbps. 

Both STUN and DTLS protocols show clear differences in their traffic patterns. For STUN, the inter packet times for Meet and Jitsi are over 9 and 4 times larger than Stadia respectively. Both conferencing apps use very infrequent packets for STUN, and this is also true for the DTLS traffic. For Stadia, we find DTLS packets every 6.38 ms, while for meet we find them every 17.4 seconds. This is the biggest difference, in that Stadia has frequent application data transmitted using DTLS, while Meet sends packets very infrequently, and for Jitsi, we only have the initial handshake of the call setting up DTLS encryption. After that, no application data is sent using DTLS. This application data for Stadia and Meet may be reports containing statistics, or information pertaining to the users involved. Which would explain its absence in Jitsi, as it does not require registering as a user to be used.

\setlength\tabcolsep{4 pt}
\renewcommand{\arraystretch}{1.2} 

	\begin{table}[t!]\centering
	\begin{small}
    \begin{tabular}{|c|c|c|c|}
  		\hline  
		\multirow{2}{*}{\textbf{Parameter}}&  	\multirow{2}{*}{\textbf{Avg. Packet  }} & 	\multirow{2}{*}{\textbf{Avg. inter } }& 	\multirow{2}{*}{\textbf{Load }}\\[5pt] 
		 &\textbf{size (bytes)} & \textbf{packet  time (ms)}& \textbf{(Mbps)}\\\hline
		
	  TR RTP  &  1159.9    &  0.88   &  10.54    \\\hline
      Meet RTP  & 1061.3    &  5.28   &   1.63   \\\hline
      Jitsi RTP  & 912.27   &  4.40   &   1.66   \\\hline	
      TR STUN  & 115.38    &  263.05    & 0.0035     \\\hline
      Meet STUN  & 134.00   &  2502.9   & 0.00042     \\\hline
      Jitsi STUN  & 118    &  1242.08   & 0.00078     \\\hline
      TR DTLS  & 148.82   & 6.38    &  0.19    \\\hline
      Meet DTLS  & 135   & 17479.01 & 0.000070      \\\hline
      jitsi DTLS  & N/A    &  N/A    & N/A      \\\hline
	\end{tabular}	
	\caption{Traffic characteristics for RTP/RTCP, DTLS and STUN streams of WebRTC applications.}
	\label{supertaula2} %
	\end{small}
	\end{table}
	
For the three applications, RTP is the main type of traffic, and it is clear that the RTP patterns are tied to the video that is being transmitted. In this regard, the main difference seems to be in terms of volume, as the load for Stadia traffic is much higher than that of the conferencing apps. The higher framerate of Stadia seems to be a main reason for the higher traffic, but this is not so clear, as the decrease in video frames (60\% and 75\% for Meet and Jitsi) is not met with an equivalent decrease in traffic load (85\% for both). Meet and Jitsi also use the same traffic load despite the fact that they operated with very different framerates, which could be a consequence of  their use of different video codecs (VP9 for Meet and VP8 for Jitsi).

To further test other video applications, we check two live streaming websites that offer live video at 1080p and 60 fps: Youtube and Twitch. We perform two captures of live streams that we can compare to our 1080p captures of Tomb Raider and check their similarities. However, neither of these two services use WebRTC. Youtube uses UDP over QUIC, and Twitch uses TCP. Figure \ref{groupingsvideo} shows the traffic patterns of all three applications, where we can observe that they operate very differently. Youtube for instance sends packets in batches every 5 seconds, which clearly shows that even though the streams are live, they do some buffering at the transmitter (adding a small latency), and in the end they operate as any other Youtube video. For Twitch the same seems to be true, but packet batches  appear much more frequently than for Youtube and they have very varied timing between batches. While live streams usually have chat interaction between content creator and viewers, this interaction does not need such strict responses as Stadia, and so both applications can use a margin of time to buffer the stream and solve typical UDP issues, such as missing packets or arrivals out of order. 

\begin{figure}[ht]
    \centering
    \includegraphics[width = 0.45\textwidth]{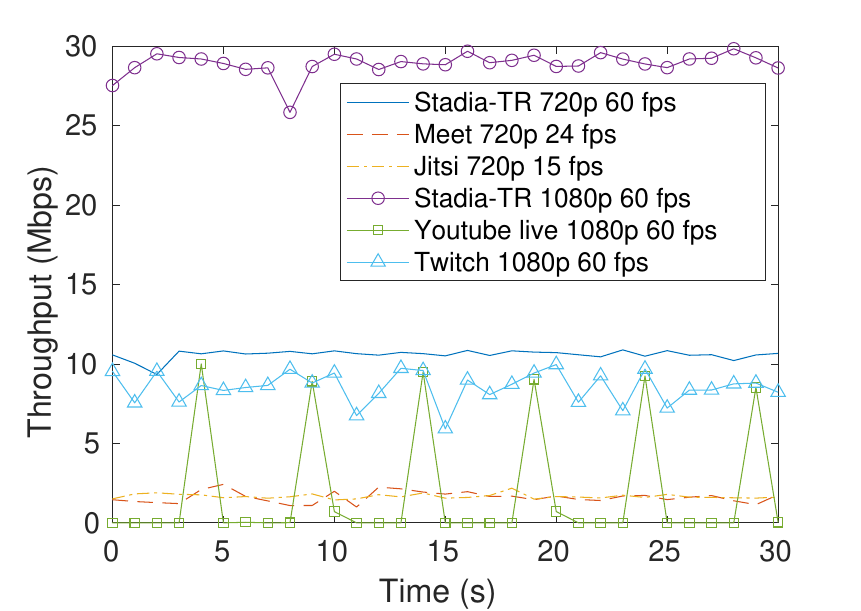}
    \caption{Throughput over 30 seconds for live streaming video applications used.}
    \label{Thrf}
\end{figure}

Finally, Figure \ref{Thrf} shows the video throughput (RTP, UDP and TCP) over 30 seconds for all the captures in this section. Here we can observe that both Twitch and Youtube, even when using 60 fps, have a much lower load than Stadia. Both live streaming applications use 10 Mbps at 1080p, close to that of a 720p Stadia stream, but less than half of a 1080p stream. Here we can also observe clearly the unique patterns of Youtube traffic, with arrivals being spaced 5 seconds, while all other services tend to use a constant stream of data.

\textit{Finding:} The traffic patterns found on Stadia streams are similar to those of other WebRTC applications. For RTP traffic the patterns are mainly a factor of video resolution and framerate, but DTLS and STUN traffic patterns seem to be application specific, as all three applications use them in different ways. The main difference in video traffic between Stadia and other WebRTC applications is the magnitude, as Stadia needs close to ten times more RTP traffic than other applications. 

Other live streaming applications such as Youtube and Twitch use different protocols and traffic patterns, even when using the same resolution and framerate, which further cements that Stadia RTP video traffic is heavily rooted in WebRTC operation, while DTLS and STUN traffic patterns are unique to Stadia.



\section{Modelling Tomb Raider's traffic}\label{sec:ModelTR}

\begin{figure}[h]
    \centering
    \includegraphics[width=0.9\columnwidth]{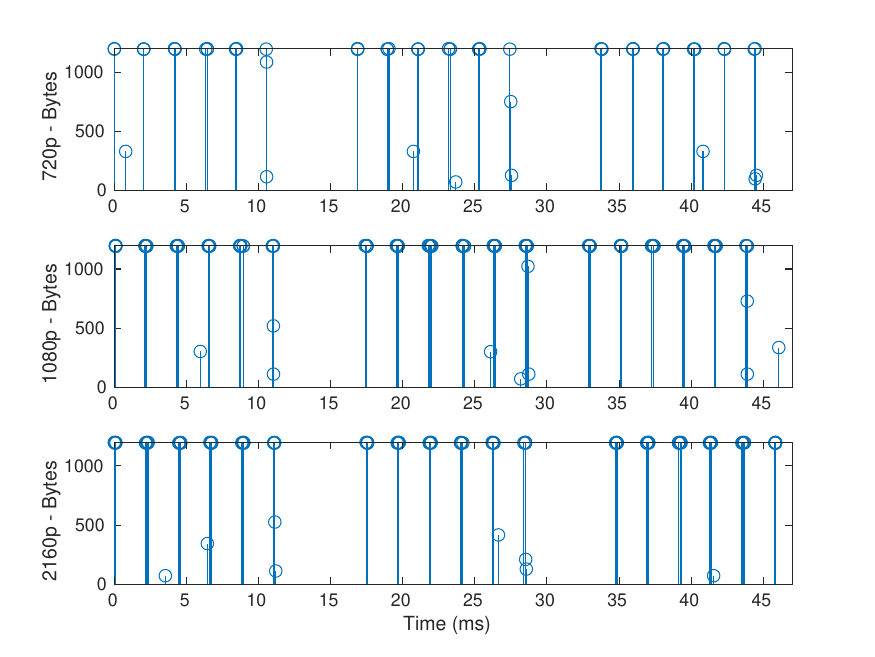}
    \caption{TR traffic patterns for different video resolutions. The y-axis indicates the size of the packets in bytes. It can be observed how the video frame timing is generally preserved, as well as the number of groups per frame. The main difference is the number of packets in each group, i.e., the batch size.} 
    \label{Fig:StadiaPatterns}
\end{figure}

\begin{figure}[h]
    \centering
    \includegraphics[width=0.8\columnwidth]{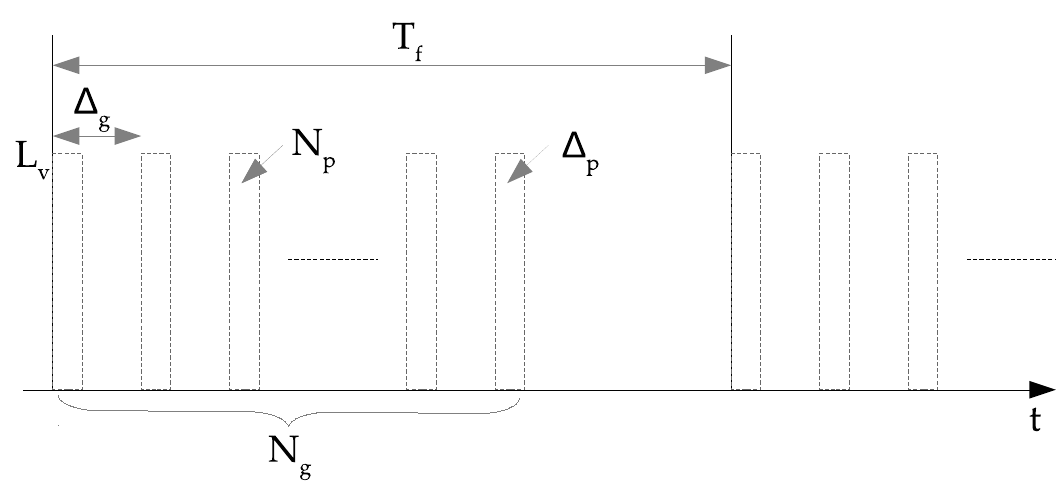}
    \caption{Representation of the temporal pattern followed by the video traffic in Stadia.}
    \label{Fig:StadiaTraffic}
\end{figure}

In this section we present a traffic model for a single Stadia game, Tomb Raider, when the VP9 codec is used, and for the three available video resolutions using the VP9 video codec: 720p, 1080p, and 2160p. 

Although there are many similarities between different Stadia games (and WebRTC apps in general, as we have seen in the previous section), the particularities of each game make it difficult to generalize, and therefore we opted to focus on a single game only, leaving for future work such a task. 

The model presented in this section has been developed by analyzing the traces from datasets D1 and D5. It takes into account the traffic patterns observed in Section~\ref{sec:und_Stadia_traffic}. It has been designed to be both accurate and simple, so it can be easily used in the performance evaluation of communication networks. 


%

The traces in dataset D1 and D5 show that regardless of the employed video resolution, TR traffic follows a clear temporal pattern that closely matches the framerate of 60 fps. In all three video resolutions (720p, 1080p, 2160p), between two video frames, we find six groups of packets in general, with an average separation of 2 ms between two groups of packets. However, the number of packets in each group depends on the video resolution. Moreover, we can also observe long ($>$1100 Bytes) and short packets, which respectively represent video (RTP) and non-video traffic (audio, STUN, and DTLS). All these aspects are shown in Figure \ref{Fig:StadiaPatterns}, where a 50 ms temporal snapshot of the Tomb Raider traffic is depicted for the three available video resolutions. 

The observed video traffic temporal patterns can be then represented as shown in Figure \ref{Fig:StadiaTraffic}, and characterized using only six parameters: the time between two frames ($T_f$), the packet size ($L_v$), the number of groups of packets per frame ($N_g$), the time between two groups of packets ($\Delta_g$), the number of packets in each group ($N_{\rm p}$), and the time between two packets inside a group ($\Delta_p$).

Regarding the non-video traffic, it can be observed in the traces that the load of non-video traffic is independent of the video resolution in use, an equal to 0.5 Mbps in all three cases. Similarly, the average packet size remains between 250 and 300 Bytes in all cases too. 

Taking those observations into account, the resulting TR traffic model is parameterized as follows. It consists of two independent streams: video and non-video traffic, that are independently modelled:
\begin{enumerate}
   \item  The common parameters for \textbf{video traffic} in all three resolutions are: Frame duration $T_f=1/60$ seconds, Video packet size $L_v=1194$ Bytes, Number of groups of packets $N_g=6$ per frame, time between groups of packets $\Delta_g=\mathcal{U}(1.5,2.5)$ ms. The time between two consecutive packets within the same group of packets is set to a constant value of $\Delta_p=0.02$ ms. Finally, depending on the video resolution, the number of packets per group ($N_{\rm pg}$) is: $N_{\rm p}=3$ for 720p, $N_{\rm p}=8$ for 1080p, and $N_{\rm p}=12$ for 2160p. 
    \item  Since the characteristics of the \textbf{non-video traffic} are independent of the video resolution, we model it as a single traffic stream of 0.5 Mbps where packet arrivals follow a Poisson process. Also, packets sizes are exponentially distributed with mean $E[L_{\rm nv}]=275$ Bytes.
\end{enumerate}

For all three video resolutions the number of packets per frame is set in a way that matches the observed traffic load in the traces (see Section \ref{sec:resolution} and Figure \ref{Fig:Figure9}). For example, for the 1080 resolution, we have 48 video packets per frame, which results in a load of $60\cdot (6\cdot 8 \cdot (1194\cdot 8))=27.5$ Mbps.


With the aim to illustrate how the presented model can be used to obtain further insights in terms of the network response in different scenarios, we extend the simulator used in \cite{bellalta2020low} to include the TR traffic model, and consider the following two examples:
\begin{enumerate}
    \item \textbf{Sharing a buffer with background traffic}: We investigate the capacity of a best-effort link with respect to the number of supported Stadia streams in presence of background traffic. The best-effort link may perfectly represent the link between the Ethernet switch and the final user as considered in the measurement campaign. The transmission rate of the link is set to $R=100$ Mbps. Background traffic is generated using a Poisson source: packet sizes are exponentially distributed with an average of $12000$ bits. The duration of each simulation is 100 s. 
    
    Figure~\ref{Fig:StadiaModel1} shows the delay (average and 99th percentile) of TR traffic with respect to the load of the background traffic when the TR traffic is generated using the presented model (model), and when it is generated directly from the traces (trace). Results are consistent since a higher background traffic load and a higher resolution, i.e., higher TR traffic load, results in higher delays. Moreover, the results obtained using the model and the ones obtained using the traces are very similar, confirming that the presented model is accurate. 
\begin{figure}[h!!!]
    \centering
    \includegraphics[width=0.90\columnwidth]{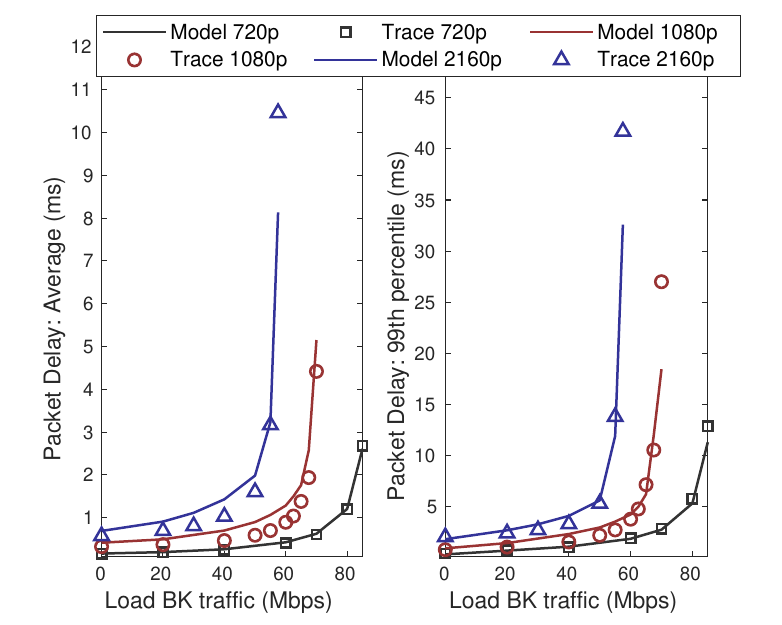}
    \caption{Average and 99th percentile delay for Tomb Raider against background traffic load. The three available video resolutions in Stadia are considered.}
    \label{Fig:StadiaModel1}
\end{figure}

    \item \textbf{Scaling the number of players}: We now consider the same 100 Mbps link. However, instead of sharing the link between Stadia and background traffic, we investigate how the latency increases when several Stadia players share the same link. To perform such an experiment, we execute as many instances of the traffic model as players. The duration of the simulation is 100 s, and each instance is initiated at a random instant of time during the first second of the simulation.
    
    Figure~\ref{Fig:StadiaModel3} shows the average and 99th percentile delay when the number of Stadia players increases, and so it does the number of traffic flows, for the three different supported video resolutions. We can observe that we can guarantee a packet delay below 3 ms, for example, in the 99\% of the cases for a single 4K player, two 1080p players, and six 720p players. Adding more users beyond those values, although supported in terms of throughput, would result in higher delays and a likely degradation of the user experience.
\begin{figure}[h!!!]
    \centering
    \includegraphics[width=0.90\columnwidth]{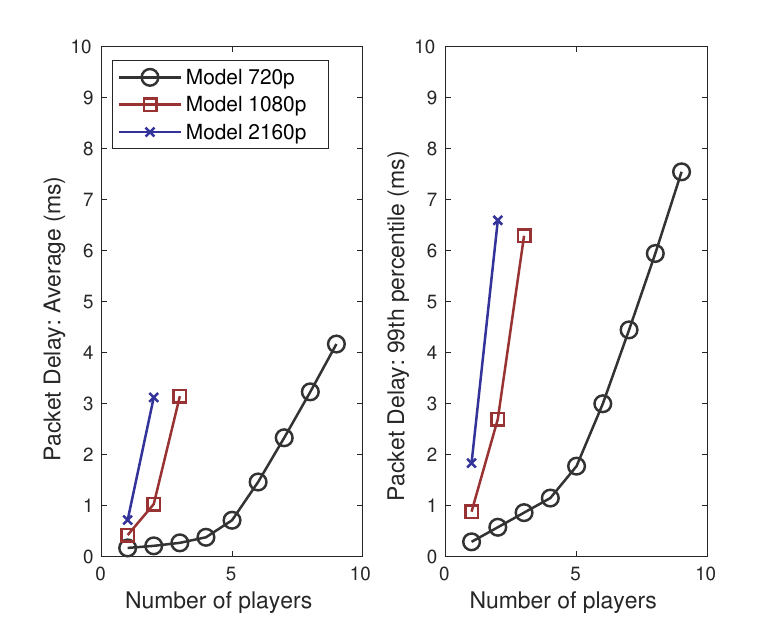}
    \caption{Increasing the number of STadia streams}
    \label{Fig:StadiaModel3}
\end{figure}

\end{enumerate}

To conclude this section, we would like to point out that the value of the model parameters can be easily adjusted to represent other traffic generation patterns, even if they are not extracted from real traces, and so it enables to investigate how a certain network reacts to different traffic generation patterns. 




\section{Conclusions}\label{sec:conclusions}

In this paper we have investigated the characteristics of the Stadia traffic. We first designed a set of experiments that implied playing some specific games under pre-defined Stadia configurations while we captured the traffic over an Ethernet network. Then, we used the collected traffic measurements to learn about the characteristics of Stadia traffic, covering from how Stadia generates the traffic at the packet level in both downlink and uplink, to how it adapts to sudden changes in the network capacity and latency. 

This paper aims to serve as a reference for future research in the area of real-time and interactive networking. In the future, it would be interesting to test Stadia performance on a Wi-Fi network, where a variety of factors can have an impact on performance, such as signal strength, number of users in the network, or the presence of other networks, as well as the particular Wi-Fi technology. Specifically, we would like to investigate how Wi-Fi is able to support low-latency in unlicensed bands \cite{adame2021time}. Other Stadia dynamics require also to be analyzed, specially in terms of its response to network latency and background traffic. Our model can also be extended, particularizing it to represent other  available games.


\section{Acknowledgements}

This work was supported by grants WINDMAL PGC2018-099959-B-I00 (MCIU/AEI/FEDER,UE), and SGR017-1188 (AGAUR).

\bibliographystyle{unsrt}

\begin{thebibliography}{10}

\bibitem{Sandvine}
Sandvine.
\newblock Global internet phenomena report.
\newblock Technical report, 2019.
\newblock
  \url{https://www.sandvine.com/hubfs/Sandvine_Redesign_2019/Downloads/Internet\%20Phenomena/Internet\%20Phenomena\%20Report\%20Q32019\%2020190910.pdf},
  Accessed on: 09 June 2020.

\bibitem{forbes}
{Ariel Shapiro}.
\newblock {Netflix Adds 15.8 Million Subscribers In First Quarter}.
\newblock
  \url{https://www.forbes.com/sites/arielshapiro/2020/04/21/netflix-stock-up-5-after-hours-reports-158-million-additional-subscribers/#5e0c10dd3d18},
  2020.
\newblock Accessed 09 June 2020.

\bibitem{yt}
{Youtube statistics for press}.
\newblock \url{https://www.youtube.com/intl/en-GB/about/press/}, 2020.
\newblock Accessed 09 June 2020.

\bibitem{twitch}
{Twitch statistics and charts}.
\newblock \url{https://twitchtracker.com/statistics}, 2020.
\newblock Accessed 09 June 2020.

\bibitem{psnow}
{How SONY Bought, And Squandered, The Future Of Gaming}.
\newblock
  \url{https://www.theverge.com/2019/12/5/20993828/sony-playstation-now-cloud-gaming-gaikai-onlive-google-stadia-25th-anniversary},
  2020.
\newblock Accessed 12 June 2020.

\bibitem{xcloud}
{Xcloud offical website}.
\newblock \url{https://www.xbox.com/en-US/xbox-game-streaming/project-xcloud},
  2020.
\newblock Accessed 12 June 2020.

\bibitem{geforce}
{Nvidia GeForce Now official website}.
\newblock \url{https://www.nvidia.com/en-us/geforce-now/ }, 2020.
\newblock Accessed 12 June 2020.

\bibitem{latencyCloudGaming}
Saeed~Shafiee Sabet, Steven Schmidt, Saman Zadtootaghaj, Babak Naderi, Carsten
  Griwodz, and Sebastian M\"{o}ller.
\newblock A latency compensation technique based on game characteristics to
  mitigate the influence of delay on cloud gaming quality of experience.
\newblock In {\em Proceedings of the 11th ACM Multimedia Systems Conference},
  MMSys ’20, page 15–25, New York, NY, USA, 2020. Association for Computing
  Machinery.

\bibitem{shea2013cloud}
Ryan Shea, Jiangchuan Liu, Edith C-H Ngai, and Yong Cui.
\newblock Cloud gaming: architecture and performance.
\newblock {\em IEEE network}, 27(4):16--21, 2013.

\bibitem{subjecti}
M.~{Jarschel}, D.~{Schlosser}, S.~{Scheuring}, and T.~{Hoßfeld}.
\newblock An evaluation of qoe in cloud gaming based on subjective tests.
\newblock In {\em 2011 Fifth International Conference on Innovative Mobile and
  Internet Services in Ubiquitous Computing}, pages 330--335, 2011.

\bibitem{modelQoE}
S.~{Zadtootaghaj}, S.~{Schmidt}, and S.~{Möller}.
\newblock Modeling gaming qoe: Towards the impact of frame rate and bit rate on
  cloud gaming.
\newblock In {\em 2018 Tenth International Conference on Quality of Multimedia
  Experience (QoMEX)}, pages 1--6, 2018.

\bibitem{singh2013performance}
Varun Singh, Albert~Abello Lozano, and Jorg Ott.
\newblock Performance analysis of receive-side real-time congestion control for
  webrtc.
\newblock In {\em 2013 20th International Packet Video Workshop}, pages 1--8.
  IEEE, 2013.

\bibitem{webrtcstudy}
Bart Jansen, Timothy Goodwin, Varun Gupta, Fernando Kuipers, and Gil Zussman.
\newblock Performance evaluation of webrtc-based video conferencing.
\newblock {\em SIGMETRICS Perform. Eval. Rev.}, 45(3):56–68, March 2018.

\bibitem{janczukowicz2016evaluation}
Ewa Janczukowicz, Arnaud Braud, St{\'e}phane Tuffin, Ahmed Bouabdallah, and
  Jean-Marie Bonnin.
\newblock Evaluation of network solutions for improving webrtc quality.
\newblock In {\em 2016 24th International Conference on Software,
  Telecommunications and Computer Networks (SoftCOM)}, pages 1--5. IEEE, 2016.

\bibitem{garcia2019understanding}
Boni Garc{\'\i}a, Micael Gallego, Francisco Gort{\'a}zar, and Antonia
  Bertolino.
\newblock Understanding and estimating quality of experience in webrtc
  applications.
\newblock {\em Computing}, 101(11):1585--1607, 2019.

\bibitem{ammar2016video}
Doreid Ammar, Katrien De~Moor, Min Xie, Markus Fiedler, and Poul Heegaard.
\newblock Video qoe killer and performance statistics in webrtc-based video
  communication.
\newblock In {\em 2016 IEEE Sixth International Conference on Communications
  and Electronics (ICCE)}, pages 429--436. IEEE, 2016.

\bibitem{bonfiglio2008detailed}
Dario Bonfiglio, Marco Mellia, Michela Meo, and Dario Rossi.
\newblock Detailed analysis of skype traffic.
\newblock {\em IEEE Transactions on Multimedia}, 11(1):117--127, 2008.

\bibitem{suznjevic2014towards}
Mirko Suznjevic, Justus Beyer, Lea Skorin-Kapov, Sebastian Moller, and Nikola
  Sorsa.
\newblock {Towards understanding the relationship between game type and network
  traffic for cloud gaming}.
\newblock In {\em {2014 IEEE International Conference on Multimedia and Expo
  Workshops (ICMEW)}}, pages 1--6. IEEE, 2014.

\bibitem{chen2013quality}
Kuan-Ta Chen, Yu-Chun Chang, Hwai-Jung Hsu, De-Yu Chen, Chun-Ying Huang, and
  Cheng-Hsin Hsu.
\newblock On the quality of service of cloud gaming systems.
\newblock {\em IEEE Transactions on Multimedia}, 16(2):480--495, 2013.

\bibitem{carrascosa2020cloudgaminganalysis}
Marc Carrascosa and Boris Bellalta.
\newblock {Cloud-gaming:Analysis of Google Stadia traffic}, 2020.

\bibitem{network1030015}
Andrea Di~Domenico, Gianluca Perna, Martino Trevisan, Luca Vassio, and Danilo
  Giordano.
\newblock {A Network Analysis on Cloud Gaming: Stadia, GeForce Now and PSNow}.
\newblock {\em Network}, 1(3):247--260, 2021.

\bibitem{9484481}
Xiaokun Xu and Mark Claypool.
\newblock {A First Look at the Network Turbulence for Google Stadia Cloud-based
  Game Streaming}.
\newblock In {\em {IEEE INFOCOM 2021 - IEEE Conference on Computer
  Communications Workshops (INFOCOM WKSHPS)}}, pages 1--5, 2021.

\bibitem{9615562}
Philippe Graff, Xavier Marchal, Thibault Cholez, Stéphane Tuffin, Bertrand
  Mathieu, and Olivier Festor.
\newblock {An Analysis of Cloud Gaming Platforms Behavior under Different
  Network Constraints}.
\newblock In {\em {2021 17th International Conference on Network and Service
  Management (CNSM)}}, pages 551--557, 2021.

\bibitem{10.1145/3491043}
Hassan Iqbal, Ayesha Khalid, and Muhammad Shahzad.
\newblock {Dissecting Cloud Gaming Performance with DECAF}.
\newblock {\em Proc. ACM Meas. Anal. Comput. Syst.}, 5(3), dec 2021.

\bibitem{10.1145/3458335.3460963}
Franck Aumont, Fr\'{e}d\'{e}rique Humbert, Christoph Neumann, Charles
  Salmon-Legagneur, and Charline Taibi.
\newblock {Dissecting Cloud Game Streaming Platforms Regarding the Impacts of
  Video Encoding and Networking Constraints on QoE}.
\newblock In {\em Proceedings of the Workshop on Game Systems (GameSys '21)},
  GameSys '21, page 13–19, New York, NY, USA, 2021. Association for Computing
  Machinery.

\bibitem{videoReport}
Bitmovin.
\newblock {Video Developer Report}.
\newblock
  \url{https://go.bitmovin.com/hubfs/Bitmovin-Video-Developer-Report-2018.pdf
  }, 2018.
\newblock Accessed 18 June 2020.

\bibitem{x265}
MultiCoreWare Inc.
\newblock {HEVC/H.265 Explained}.
\newblock \url{http://x265.org/hevc-h265/ }, 2020.
\newblock Accessed 18 June 2020.

\bibitem{x265_2}
ITU-T.
\newblock {Joint Collaborative Team on Video Coding}.
\newblock
  \url{https://www.itu.int/en/ITU-T/studygroups/2017-2020/16/Pages/video/jctvc.aspx
  }, 2020.
\newblock Accessed 18 June 2020.

\bibitem{h265Comp}
J.~{Bienik}, M.~{Uhrina}, M.~{Kuba}, and M.~{Vaculik}.
\newblock {Performance of H.264, H.265, VP8 and VP9 Compression Standards for
  High Resolutions}.
\newblock In {\em {2016 19th International Conference on Network-Based
  Information Systems (NBiS)}}, pages 246--252, 2016.

\bibitem{h265VP9}
T.~{Uhl}, J.~H. {Klink}, K.~{Nowicki}, and C.~{Hoppe}.
\newblock {Comparison Study of H.264/AVC, H.265/HEVC and VP9-Coded Video
  Streams for the Service IPTV}.
\newblock In {\em {2018 26th International Conference on Software,
  Telecommunications and Computer Networks (SoftCOM)}}, pages 1--6, 2018.

\bibitem{chromium}
Google.
\newblock {Chromium Blog: Celebrating 10 years of WebM and WebRTC}.
\newblock
  \url{https://blog.chromium.org/2020/05/celebrating-10-years-of-webm-and-webrtc.html
  }, 2020.
\newblock Accessed 18 June 2020.

\bibitem{opusA}
Opus.
\newblock {Opus audio codec }.
\newblock \url{https://opus-codec.org/ }, 2020.
\newblock Accessed 06 August 2020.

\bibitem{icerfc}
{Interactive Connectivity Establishment (ICE): A Protocol for Network Address
  Translator (NAT) Traversal}.
\newblock \url{https://datatracker.ietf.org/doc/html/rfc8445}, 2018.
\newblock Accessed 29 August 2021.

\bibitem{dtlsrfc}
{Datagram Transport Layer Security Version 1.2}.
\newblock \url{https://datatracker.ietf.org/doc/html/rfc6347}, 2018.
\newblock Accessed 29 August 2021.

\bibitem{srtp}
{Datagram Transport Layer Security (DTLS) Extension to Establish Keys for the
  Secure Real-time Transport Protocol (SRTP)}.
\newblock \url{https://datatracker.ietf.org/doc/html/rfc5764}, 2010.
\newblock Accessed 29 August 2021.

\bibitem{rtprfc}
{RTP: A Transport Protocol for Real-Time Applications}.
\newblock \url{https://datatracker.ietf.org/doc/html/rfc3550}, 2003.
\newblock Accessed 29 August 2021.

\bibitem{rtpext}
{Extended RTP Profile for Real-time Transport Control Protocol (RTCP)-Based
  Feedback (RTP/AVPF)}.
\newblock \url{https://datatracker.ietf.org/doc/html/rfc4585}, 2006.
\newblock Accessed 29 August 2021.

\bibitem{gcc}
Gaetano Carlucci, Luca De~Cicco, Stefan Holmer, and Saverio Mascolo.
\newblock Analysis and design of the google congestion control for web
  real-time communication (webrtc).
\newblock In {\em Proceedings of the 7th International Conference on Multimedia
  Systems}, MMSys ’16, New York, NY, USA, 2016. Association for Computing
  Machinery.

\bibitem{gNegL}
TechRepublic.
\newblock {Google Stadia's biggest challenge with streaming and meeting gamers'
  expectations }.
\newblock
  \url{https://www.techrepublic.com/article/google-stadias-biggest-challenge-with-steaming-and-meeting-gamers-expectations/
  }, 2020.
\newblock Accessed 27 July 2020.

\bibitem{Stadiainfra}
IEEE~spectrum Jeremy~Hsu.
\newblock {How YouTube Paved the Way for Google's Stadia Cloud Gaming Service}.
\newblock
  \url{https://spectrum.ieee.org/tech-talk/telecom/internet/how-the-youtube-era-made-cloud-gaming-possible},
  2020.
\newblock Accessed 06 August 2020.

\bibitem{outatime}
Kyungmin Lee, David Chu, Eduardo Cuervo, Johannes Kopf, Yury Degtyarev, Sergey
  Grizan, Alec Wolman, and Jason Flinn.
\newblock Outatime: Using speculation to enable low-latency continuous
  interaction for mobile cloud gaming.
\newblock In {\em Proceedings of the 13th Annual International Conference on
  Mobile Systems, Applications, and Services}, MobiSys ’15, page 151–165,
  New York, NY, USA, 2015. Association for Computing Machinery.

\bibitem{tanks}
Mark Claypool and Kajal Claypool.
\newblock Latency can kill: Precision and deadline in online games.
\newblock In {\em Proceedings of the First Annual ACM SIGMM Conference on
  Multimedia Systems}, MMSys ’10, page 215–222, New York, NY, USA, 2010.
  Association for Computing Machinery.

\bibitem{googleRec}
Google.
\newblock {Bandwidth, data usage, and stream quality }.
\newblock \url{https://support.google.com/stadia/answer/9607891?hl=en }, 2020.
\newblock Accessed 19 June 2020.

\bibitem{bellalta2020low}
Boris Bellalta.
\newblock On the low-latency region of best-effort links for delay-sensitive
  streaming traffic.
\newblock {\em IEEE Communications Letters}, 25(3):970--974, 2020.

\bibitem{adame2021time}
Toni Adame, Marc Carrascosa-Zamacois, and Boris Bellalta.
\newblock {Time-sensitive networking in IEEE 802.11 be: On the way to
  low-latency Wi-Fi 7}.
\newblock {\em Sensors}, 21(15):4954, 2021.

\end{thebibliography}

\end{document}